\documentclass[%
 reprint,
nofootinbib,
 amsmath,amssymb,
 aps,
 prd
]{revtex4-2}
\usepackage{graphicx}
\usepackage{dcolumn}
\usepackage{bm}

\usepackage{latexsym,verbatim}
\usepackage{epsfig,epstopdf,color}
\usepackage{float}
\usepackage{soul}
\usepackage{physics}
\usepackage{esint}
\usepackage{empheq}
\usepackage[normalem]{ulem}
\usepackage{xcolor}
\usepackage{amsthm}
\usepackage{bbm}
\usepackage{mathrsfs}
\newcommand{\stkout}[1]{\ifmmode\text{\sout{\ensuremath{#1}}}\else\sout{#1}\fi}

\usepackage[colorlinks=true, allcolors=blue]{hyperref}

\usepackage{listings}
\usepackage{soul}

\newcommand{\Lie}{\pounds}

\newcommand{\DC}[1]{\textcolor{red}{[DC: #1]}}
\newcommand{\ny}[1]{\textcolor{blue}{{{[NY: #1]}}}}

\begin{document}

\title{Neutron Star Radial Perturbations for Causal, Viscous, Relativistic Fluids}
\author{Daniel A. Caballero}
\affiliation{Illinois Center for Advanced Studies of the Universe \& Department of Physics,
University of Illinois at Urbana-Champaign, Urbana, Illinois 61801, USA.}

\author{Nicol\'as Yunes}
\affiliation{Illinois Center for Advanced Studies of the Universe \& Department of Physics,
University of Illinois at Urbana-Champaign, Urbana, Illinois 61801, USA.}

\begin{abstract}
    Which of the multiple models of causal and stable relativistic viscous fluids that have been developed is best suited to describe neutron stars?
    The modeling of out-of-equilibrium effects in these relativistic, astrophysical objects must be done with care, as simple Newtonian intuition fails to remain causal. 
    Radial stability of neutron stars is one of the primary conditions for the viability of such out-of-equilibrium models. 
    In this paper, we study radial perturbations of neutron stars for the Eckart, the Bemfica-Disconzi-Noronha-Kovtun, and the M\"uller-Israel-Stewart fluid models of relativistic viscous fluids.
    We find that for small viscosity, the three models have the same stability properties: they are always stable to bulk and shear viscosity, but they can be unstable to heat conductivity if certain thermodynamic conditions are violated.
    For the latter case, we derive a necessary criterion for stability to heat conductivity that applies to all three fluids.
    Moreover, we show that the additional degrees of freedom introduced by the Bemfica-Disconzi-Noronha-Kovtun and the M\"uller-Israel-Stewart models force the perturbations to evolve on fast timescales.
    Specifically, the Bemfica-Disconzi-Noronha-Kovtun model has additional oscillatory perturbations that propagate with the speed of second sound, while the M\"uller-Israel-Stewart model MIS only exhibits decaying behavior on the fast timescale.
    This work therefore establishes the first formal results and criteria for radial stability of these three out-of-equiblirium fluid models on the non-trivial, relativistic background of neutron stars.
%
\end{abstract}

\maketitle

\allowdisplaybreaks[4]

\section{Introduction}

Determining the behavior of out-of-equilibrium relativistic fluids has been a contested question for decades.
A well-behaved theory of relativistic fluids must satisfy a number of conditions:
(i) causality, i.e.~the propagation speed of the fluid must be finite and less than the speed of light;
(ii) stability, i.e.~plane wave perturbations about a thermodynamic equilibrium in flat spacetime must be linearly stable;
and (iii) local well-posedness, i.e.~there must exist solutions to the initial value problem and such solutions must be unique.
Multiple theories developed to model relativistic viscous fluids\footnote{We will use the word viscous to refer to all out-of-equilibrium effects and not just viscosity, e.g.~heat conductivity.} have been shown to be ill-behaved;
for example, the Eckart and Landau-Lifshitz fluid models have been shown to violate both causality and stability in relativistic settings \cite{Pichon1965,Hiscock1985,RezzollaZanotti2013}.

In recent years, the Bemfica-Disconzi-Noronha-Kovtun (BDNK) fluid model has been developed to address these issues \cite{Bemfica2018,Bemfica2019,Hoult2020,Bemfica2022}.
The BDNK ``theory'' is based on the idea that all models of relativistic fluids are effective field theories based on a gradient expansion of the variables of motion \cite{Kovtun2019}.
The BDNK model, of course, is not the \textit{only} model that satisfies these properties,
but it is the first one shown to satisfy causality in the nonlinear regime and stability (about a flat spacetime background), while remaining of first order in the gradient expansion \cite{Bemfica2022}.
In this paper, we will mostly focus on the BDNK fluid model because it has the strongest, proven results \cite{Bemfica2022,Disconzi2024}, and we will compare it with the Eckart \cite{Eckart1940} and the M\"uler-Israel-Stewart (MIS) fluid models \cite{Israel1976,IsraelStewart1979}, the latter of which has been shown to be causal and stable at least in the linear regime.

Neutron stars give us one of the best ways to probe the out-of-equilibrium effects of fluids in the relativistic regime.
The strong gravitational fields inside neutron stars must be treated in a relativistic setting, and thus, any viscous effect must be modeled using theories of relativistic viscous fluids.
Weak-force processes interior to neutron stars are expected to lead to viscous effects \cite{Alford2018}.
The most relevant effect is expected to be bulk viscosity, which emerges from interactions between protons and neutrons returning to $\beta$-equilibrium \cite{Haensel1992,Gavassino2021,Most2022,Yang2023},
and from interactions between hyperons that may exist inside the neutron star core \cite{LindblomOwen2002,Saketh2024,AlfordHaber2021}.
Shear viscosity is also expected to arise inside neutron stars from electron-electron scattering \cite{Alford2018,Shternin2008},
but the coefficient of shear viscosity is expected to be smaller than that of bulk viscosity \cite{Sawyer1989}.
These viscous effects are expected to have measurable effects in the gravitational-wave signal of neutron stars,
in both the inspiral \cite{Most2022,Abhishek2024,Ripley2023} and the post-merger stages \cite{Most2024,Chabanov2025,Espino2023}, at least for a short period of time.

Whether neutron stars are stable to out-of-equilibrium effects remains an open question.
Newtonian, cold, rotating, compact stars (e.g.~white dwarfs) have a viscous instability \cite{Roberts1963,Ellipsoidal},
but it is counteracted by the Chandrasekhar-Friedman-Schutz (CFS) instability\footnote{The CFS instability is generic to all rotating compact stars, and it occurs due to gravitational radiation. See \cite{Chandrasekhar1970,FriedmanSchutz1978b} for the Newtonian problem and \cite{FriedmanSchutz1975,Friedman1978} for the relativistic problem.}, so that viscosity can in fact stabilize the star \cite{Chandrasekhar1970,Lindblom1983}.
A similar result was later established in full General Relativity (GR) for the Eckart model \cite{Lindblom1983}, but as mentioned earlier, this model is not well-behaved.
Further work in this area was carried out by \cite{Cutler1987}, who calculated the damping time of modes due to shear viscosity for an Eckart fluid, and \cite{Camelio2023} who simulated spherically-symmetric neutron stars for MIS-type fluids.
The very recent work of \cite{RedondoYuste2024} examined non-radial perturbations of neutron stars for a BDNK fluid without heat conductivity, but it did not consider stability conditions.
Perturbations of BDNK neutron stars have also been used to study the scattering of gravitational waves off viscous neutron stars \cite{Boyanov2024,RedondoYuste2025}, which can lead to superradiance in the slowly rotating case, but such superradiance does not induce an instability \cite{RedondoYuste2025}.

In this paper, we examine whether cold, non-rotating neutron stars are radially stable in full GR when including out-of-equilibrium effects through Eckart, BDNK, and MIS fluid models. This work differs from the previous results summarized above because we focus on and more formally prove radial stability, we consider only radial perturbations (the $l=0$ case), and we include heat conductivity effects, as well as bulk and shear viscosity. Although the study we conduct in this paper is limited to radial stability, the latter is a necessary condition for the viability of any astrophysical object. Moreover, we expect that the results we find when we study radial stability will give us insight for the general stability problem of non-radial modes.

We first determine the corrections to the eigenvalues of a neutron star in equilibrium for small viscosities. We make this approximation because astrophysical compact objects are expected to have small out-of-equilibrium effects, even when they merge with other compact objects~\cite{Most2022,Most2024}. Working to first-order in small viscosities, we derive conditions for a neutron star to be radially stable and find that they are the same for all three fluid models. In particular, we discover that neutron stars are always radially stable to bulk and shear viscosities. We also discover that heat conductivity effects can render neutron stars unstable unless certain conditions are satisfied. In particular, we derive a necessary and sufficient integral condition for radial stability to heat-conductivity effects, as well as simpler, necessary integral condition for high-frequency radial stability. From the latter, we then derive two sufficient algebraic conditions for high-frequency radial stability that depend on the speed of sound of the fluid and the derivative of the pressure with respect to energy density while holding baryon number constant. 

While studying stability, we also find that radial perturbations of BDNK and MIS fluids depend on two time scales (a fast one and a slow one), which separate when viscosity is small. Multiple-scale analysis allows us to prove that radial perturbations of MIS fluids present exponentially decaying behavior on the fast timescale, while BDNK fluids also have additional oscillatory behavior that also evolves on the fast timescale. To our knowledge, this is the first work to introduce a fast timescale in the evolution of perturbations within the context of relativistic viscous theories. The above result yields a phenomenological difference between the evolution of radial perturbations of these two fluid models. However, the perturbations decay away exponentially fast for small viscosity, irrespective of the fluid model analyzed, making the difference difficult to measure in practice.

We emphasize that the stability of a given hydrodynamic theory (which had been shown e.g.~in the BDNK case already in~\cite{Bemfica2022}) is not the same as the stability of neutron stars within the same fluid model. The condition of stability that BDNK and other models satisfy, refers to \textit{plane wave} solutions that perturb the thermodynamic equilibrium in \textit{flat spacetime}  \cite{Disconzi2024,Bemfica2022}. Therefore, the stability of neutron stars differs from the stability of the theory in two ways:
(i) neutron stars have strong gravitational fields, and thus spacetime is not flat, and we must consider couplings to the gravitational sector; and
(ii) the boundary of a neutron star is finite, while plane wave solutions assume that the domain is infinite.
Therefore, stability of a hydrodynamic theory and stability of a neutron star in said hydrodynamic theory do not necessarily imply each other because they are concerned with the stability of different systems under the same model. In light of this, our work now enables the complete study of stability of neutron stars, beyond radial modes, in relativistic viscous theories.  

The remainder of this paper presents the details of the analysis that yielded the results summarized above, and it is organized as follows. 
In Sec.~\ref{sec:theo}, we provide a brief overview on the equations of relativistic viscous hydrodynamics and discuss the fluid models that we will be examining in this paper, namely the Eckart, BDNK, and MIS (in its Maxwell-Cattaneo form) fluid models.
In Sec.~\ref{sec:perts}, we examine the treatment of Lagrangian perturbations for a general theory of relativistic out-of-equilibrium fluids.
In Sec.~\ref{sec:spher}, we consider the radial perturbation for each of the three fluid models individually and derive equations of motion for the radial perturbations for each fluid.
We then evaluate these resulting equations for small viscosity in Sec.~\ref{sec:small} up to first order, and determine stability conditions for the different fluid models.
Finally, we conclude in Sec.~\ref{sec:conclusions}.
We use the following conventions in this paper: Greek letters indices stand for spacetime coordinates, the metric signature is $(-,+,+,+)$, and we use geometric units where $G=c=k_B=1$.


\section{Relativistic Viscous Theories}\label{sec:theo}

In this section, we start by reviewing some basics of hydrodynamics and relativistic viscous fluids.
We also introduce the specific models we examine in Sec.~\ref{ssec:rela}, following mostly \cite{RezzollaZanotti2013}.

\subsection{Relativistic Viscous Hydrodynamics}

We model neutron stars as solutions to the Einstein field equations coupled to matter through the stress-energy tensor of the fluid, namely
\begin{subequations}\label{eq:eom}
    \begin{align}
        G_{\mu\nu}&=8\pi T_{\mu\nu} \ ,\\
        \nabla_\nu T^{\mu\nu}&=0 \ . \label{eq:euler}
    \end{align}
\end{subequations}
Moreover, we generally include a conserved current $J^\mu$, which we interpret as the baryon number current and is also conserved, i.e.~it satisfies
\begin{equation}\label{eq:consb}
    \nabla_\mu J^\mu=0 \ .
\end{equation}
The stress-energy tensor is usually taken to be that of a perfect fluid, that is
\begin{equation}
    T_{\rm PF}^{\mu\nu}=\varepsilon \, u^\mu u^\nu+p\Pi^{\mu\nu} \ ,
\end{equation}
where $\varepsilon$ is the energy density of the fluid, $p$ its pressure, $u^\mu$ its four velocity -- normalized such that $u_\mu u^\mu=-1$ -- and $\Pi^{\mu\nu}=u^\mu u^\nu+g^{\mu\nu}$ is the projector perpendicular to $u^\mu$.
The conserved current also has a perfect-fluid form as follows
\begin{equation}
    J_{\rm PF}^\mu=nu^\mu \ ,
\end{equation}
where $n$ is the baryon number density. 
To close the system, we must include an equation of state, which we write as $[p(n,s),\varepsilon(n,s)]$ where $s$ is the entropy per baryon, and is related to the other quantities via the first law of thermodynamics
\begin{equation}
    d\varepsilon=\frac{\varepsilon+p}{n}dn+nTds \ .
\end{equation}

To obtain the equations of motions for a viscous fluid, we add additional terms to the stress-energy tensor and baryon number current
\begin{subequations}\label{eq:corr}
    \begin{align}
        T^{\mu\nu}&=T_{\rm PF}^{\mu\nu}+T_{\rm NPF}^{\mu\nu} \ , \\
        J^\mu&=J_{\rm PF}^\mu+J_{\rm NPF}^{\mu} \ .
    \end{align}
\end{subequations}
The non-perfect contributions to the stress-energy tensor and the baryon number current can be decomposed into 
\begin{subequations}
    \begin{align}
        T_{\rm NPF}^{\mu\nu}&=\mathscr E u^\mu u^\nu +\mathscr P\Pi^{\mu\nu}+u^\mu\mathscr Q^\nu+u^\nu\mathscr Q^\mu +\pi^{\mu\nu} \ ,\\
        J_{\rm NPF}^\mu&=\mathscr N u^\mu+\mathscr J^\mu \ ,
    \end{align}
\end{subequations}
with 
\begin{subequations}\label{eq:flux}
    \begin{align}
        \mathscr E&=u_\mu u_\nu T_{\rm NPF}^{\mu\nu} \ ,\\
        \mathscr P&=\frac13\Pi_{\mu\nu}T_{\rm NPF}^{\mu\nu} \ ,\\
        \mathscr Q^\mu&=-\Pi^\mu_\lambda u_\nu T_{\rm NPF}^{\lambda\nu} \ ,\\
        \pi^{\mu\nu}&=\Pi^\mu_\alpha\Pi^\nu_\beta T_{\rm NPF}^{\alpha\beta}-\frac13(\Pi_{\alpha\beta}T_{\rm NPF}^{\alpha\beta})\Pi^{\mu\nu} \ ,\\
        \mathscr N &= -u_\mu J_{\rm NPF}^\mu \ ,\\
        \mathscr J^\mu&=\Pi^{\mu}_\nu J_{\rm NPF}^\nu \ .
    \end{align}
\end{subequations}
The quantities $\mathscr E,\mathscr P,\mathscr Q^\mu,\pi^{\mu\nu},\mathscr N$, and $\mathscr J^\mu$ are called the viscous fluxes or corrections,
and this is a general decomposition for any symmetric tensor $T^{\mu\nu}$ and vector $J^\mu$ \cite{Kovtun2019}.
To solve for the system Eqs.~\eqref{eq:eom}-\eqref{eq:consb}, we still require specifying the viscous fluxes as functions of $u^\mu,n$, and $s$.
Any such specification is known as a constitutive relation, and each different relation results in a different viscous hydrodynamic fluid model, or ``theory'' for short.

Before we specify the relations for the BDNK model that we analyze in this paper,
let us write what the equations of motion are for a general theory.
Generally, instead of dealing with Eq.~\eqref{eq:euler}, we decompose the equation into a parallel component $u_\mu \nabla_\nu T^{\mu\nu}$ and a perpendicular one $\Pi^\sigma_\mu\nabla_\nu T^{\mu\nu}$.
For the stress-energy tensor and baryon number current from Eq.~\eqref{eq:corr}, the resulting energy-momentum conservation equations are \cite{RezzollaZanotti2013,Bemfica2022}
\begin{align}
    u^\mu\nabla_\mu\varepsilon+(\varepsilon+p)\Theta&=-u^\mu\nabla_\mu \mathscr E-(\mathscr E+\mathscr P)\Theta \nonumber\\
    &\ -\mathscr Q_\mu a^\mu-\nabla_\mu \mathscr Q^\mu -\pi^{\mu\nu}\sigma_{\mu\nu}\ , \label{eq:para}\\
    (\varepsilon+p)a^\mu+\Pi^{\mu\lambda}\nabla_\lambda p&= -(\mathscr E+\mathscr P)a^\mu-\Pi^{\mu\nu}\nabla_\nu\mathscr P \nonumber\\
    &\ -\Pi^{\mu\lambda}\nabla_\nu\pi^\nu_\lambda-u^\nu\nabla_\nu\mathscr Q^\mu-\frac43\mathscr Q^\mu\Theta \nonumber\\
    &\ -\mathscr Q^\nu[\sigma_\nu^\mu+\omega_{\nu}^{\ \mu}]\ , \label{eq:perp}\\
    n\Theta+u^\mu\nabla_\mu n&=-\mathscr N\Theta-u^\mu\nabla_\mu\mathscr N-\nabla_\mu\mathscr J^\mu \ . \label{eq:bary}
\end{align}
where the quantities $\Theta, a^\mu,\sigma_{\mu\nu}$, and $\omega_{\mu\nu}$ come from the decomposition of the derivative of the four velocity, i.e.,
\begin{subequations}\label{eq:deru}
    \begin{align}
        \Theta&\equiv \nabla_\mu u^\mu \ ,\\
        a^\mu&\equiv u^\nu\nabla_\nu u^\mu \ ,\\
        \sigma_{\mu\nu}&\equiv\Pi^\alpha_\mu \Pi^\beta_\nu(\nabla_{(\alpha}u_{\beta)}-\tfrac13g_{\alpha\beta}\Theta) \ ,\\
        \omega_{\mu\nu}&\equiv\Pi^\alpha_\mu \Pi^\beta_\nu (\nabla_{[\alpha}u_{\beta]})\ .
    \end{align}
\end{subequations}

\subsection{Constitutive Relations}\label{ssec:rela}

We now examine the constitutive relations [i.e., specification of the viscous fluxes of Eq.~\eqref{eq:flux}] that we will analyze in this paper.
We will mainly focus on the BDNK fluid model where the viscous fluxes are given by \cite{Bemfica2022}
\begin{subequations}\label{eq:bdnk}
    \begin{align}
        \mathscr E_{BDNK}&=\tau_{\mathscr E}\left[u^\mu\nabla_\mu \varepsilon+(\varepsilon+p)\Theta\right] \ , \label{eq:bdnke}\\
        \mathscr P_{BDNK}&=-\zeta\Theta+\tau_{\mathscr P}\left[u^\mu\nabla_\mu \varepsilon+(\varepsilon+p)\Theta\right] \ , \label{eq:bdnkp}\\
        \mathscr Q^\mu_{BDNK}&=\kappa T\frac{\varepsilon+p}{n}\Pi^{\mu\lambda}\nabla_\lambda\varphi \nonumber\\
        &\quad +\tau_{\mathscr Q}\left[(\varepsilon+p)a^\mu+\Pi^{\mu\lambda}\nabla_\lambda p\right] \ , \label{eq:bdnkq}\\
        \pi^{\mu\nu}_{ BDNK}&=-2\eta\sigma^{\mu\nu} \ ,\\
        \mathscr N_{BDNK}&=0 \ ,\quad \mathscr J^\mu_{BDNK}=0 \ .
    \end{align}
\end{subequations}
where $\varphi$ is known as the fugacity, is constant in equilibrium and is given by \cite{RezzollaZanotti2013,Hiscock1985}
\begin{equation}
    \varphi\equiv \frac{\varepsilon+p}{nT}-s \ .
\end{equation}
Meanwhile, the variables $\zeta,\eta,\kappa$, and $\tau_{\mathscr E,\mathscr P,\mathscr Q}$ are known as transport coefficients that must also be specified in terms of the thermodynamic quantities (which in our case are $n$ and $s$) to describe the theory.
These dependencies should be determined via microphysics model of the interactions.

In this paper, we will also compare the BDNK results with those we obtain for an Eckart fluid, whose constitutive relations are given by \cite{Eckart1940,RezzollaZanotti2013}
\begin{subequations}\label{eq:Eckart}
    \begin{align}
        \mathscr P_{Eck}&=-\zeta\Theta \ , \label{eq:eckp}\\
        \mathscr Q^\mu_{Eck}&=\kappa T\frac{\varepsilon+p}{n}\Pi^{\mu\lambda}\nabla_\lambda \varphi \ , \label{eq:eckq}\\
        \pi^{\mu\nu}_{Eck}&=-2\eta\sigma^{\mu\nu} \ , \label{eq:eckpi}\\
        \mathscr E_{Eck}&=0 \ ,\quad \mathscr N_{Eck}=0 \ ,\quad \mathscr J_{Eck}^\mu=0 \ .
    \end{align}
\end{subequations}
We note that the original prescription by Eckart uses a different heat current, given by \cite{Eckart1940}
\begin{equation}\label{eq:oeck}
    \mathscr Q^\mu_{OEck}=-\kappa T\left[\Pi^{\mu\lambda}\nabla_\lambda\log T+a^\mu\right] \ ;
\end{equation}
however, the two heat fluxes [from Eqs.~\eqref{eq:eckq} and \eqref{eq:oeck}] are equivalent in a first-order theory up to field redefinitions \cite{Kovtun2019}.
Nevertheless, the stability properties (in the sense of stability of traveling waves in flat spacetime) of the two heat fluxes differ, with modes unstable with heat flux Eq.~\eqref{eq:oeck} becoming stable when using the heat flux Eq.~\eqref{eq:eckq} \cite{Kovtun2012}.
We use the heat flux of Eq.~\eqref{eq:eckq} due to it being closer to the BDNK fluid model (notice how Eq.~\eqref{eq:Eckart} is obtained from Eq.~\eqref{eq:bdnk} by taking $\tau_{\mathscr E},\tau_{\mathscr P},\tau_{\mathscr Q}\to 0$) and we will refer to the resulting theory as the Eckart model.
We note that a heat flux of the form Eq.~\eqref{eq:eckq} has previously been used in the study of modes of neutron stars in \cite{Cutler1987}.

Both, the Eckart and BDNK fluid models are within a wider class of relativistic viscous theories known as first-order theories,
where the constitutive relations are given up to first-order in derivatives of the four velocity and the thermodynamic quantities.
Previous work \cite{Kovtun2019} has looked at the most general possible first-order theory, which results in $14$ transport coefficients.
However, using field redefinitions, one can reduce the system to just $6$ transport coefficients.
The BDNK model is one such way to redefine the fields, which has been shown to be causal, stable (in the sense of wave stability in flat spacetime) and strongly hyperbolic.
One must note that when the theory is referred to as causal, stable and strongly hyperbolic, it rather means that these properties hold under some non-empty conditions,
relating the transport coefficients to each other \cite{Bemfica2022}.
For the purpose of this work, we can assume that we have already chosen such coefficients, so that the theory is causal and stable.

Another class of theories are second-order ones, sometimes also known as Israel-Stewart-type theories, where the constitutive relations are instead given in terms of up to second derivatives of the four velocity and the thermodynamic variables.
These were first developed to solve the causality and stability issues known to occur for the Eckart and other first-order theories of relativistic viscous fluids \cite{Israel1976,IsraelStewart1979}.
In this paper, we compare our BDNK and Eckart results with those we obtain for the Israel-Stewart fluid model in its Maxwell-Cattaneo form\footnote{Once again, the original theory has a different constitutive relation in which the right-hand side of Eq.~\eqref{eq:mcq} is given by the right-hand side of  Eq.~\eqref{eq:oeck}. However, we use here a modified version of the Israel-Stewart equations, such that the right-hand side is given by Eq.~\eqref{eq:mcq} instead, and the equations are more similar to the BDNK ones.} \cite{RezzollaZanotti2013}
\begin{subequations}\label{eq:mc}
    \begin{align}
    \tau_0u^\nu\nabla_\nu\mathscr P_{MC}+\mathscr P_{MC}&=-\zeta\Theta \ ,\\
    \tau_1\Pi^{\mu}_{\nu}u^\lambda\nabla_\lambda\mathscr Q^\nu_{MC}+\mathscr Q_{MC}^\mu&=\kappa T\frac{\varepsilon+p}{n}\Pi^{\mu\lambda}\nabla_\lambda\varphi \ , \label{eq:mcq}\\
    \tau_2\Pi_\alpha^\mu\Pi_\beta^\nu u^\lambda\nabla_\lambda\pi_{MC}^{\alpha\beta}+\pi_{MC}^{\mu\nu}&=-2\eta\sigma^{\mu\nu} \ ,\\
    \mathscr E_{MC}=0 \ ,\quad \mathscr N_{MC}&=0 \ , \quad \mathscr J^\mu_{MC}=0 \ .
    \end{align}
\end{subequations}
We use the Maxwell-Cattaneo form instead of the full equations because the additional terms will be of higher order in perturbations for the analysis we carry out in this paper.
We note that there are more second-order theories that have been shown to be causal and stable, but, for brevity, we focus on the Israel-Stewart fluid model due to its simplicity.
Moreover, in the linearization of the equations, many of these theories result in similar equations,
e.g.~the Denicol-Niemi-Moln\'ar-Rischke (DNMR) theory \cite{Denicol2012} when considering no corrections to the baryon number current leads to the same equations in the linear regime as the Maxwell-Cattaneo model with no heat flux.
For a more in-depth review of the multiple models of relativistic viscous fluids, we refer the reader to \cite{Rocha2024}.

\section{General Perturbations Equations}\label{sec:perts}

We now derive the perturbation equations for viscous fluids.
We carry this out through the use of Lagrangian perturbations, similar to what is done in \cite{Friedman1978} for a perfect fluid.
We differentiate between the Eulerian variation, denoted by $\delta$, and the Lagrangian variation, denoted by $\Delta$.
In our notation, any given quantity that is \textit{not} preceded by either $\delta$ or $\Delta$ refers to a \textit{background} variable. For example, if the quantity under consideration is the pressure, then the symbol $p$ refers to the background pressure, while $\delta p$ is its Eulerian variation and $\Delta p$ its Lagrangian variation.

The Lagrangian displacement $\xi^\mu$ is the generator of the perturbations, and for any (tensor) quantity $T$, its Eulerian and Lagrangian variations are related by
\begin{equation}\label{eq:lie}
    \Delta T=\delta T+\Lie_\xi T \ ,
\end{equation}
where $\Lie_\xi T$ is the Lie derivative along $\xi^\mu$ of the \textit{background} quantity $T$.
This means that for any quantity that is zero in the background, its Eulerian and Lagrangian variation are the same (i.e.~if $T=0$ then $\Delta T=\delta T$).

As shown in \cite{Friedman1978}, the expression for the Lagrangian variation of the four velocity is\footnote{ Equation~\eqref{eq:delumu} involves \textit{Lagrangian} perturbations, therefore large Eulerian perturbations in the four velocity $\delta u^\mu$ can result even if the Eulerian perturbation of the metric is negligible (i.e. even if $h_{\mu\nu}=0$, one obtains $\delta u^\mu\ne 0$).}
\begin{equation}\label{eq:delumu}
    \Delta u^\mu=\frac12 u^\mu u^\alpha u^\beta \Delta g_{\alpha\beta} \ ,
\end{equation}
where the Lagrangian variation of the metric can be related to the Eulerian variation of the metric $\delta g_{\mu\nu}\equiv h_{\mu\nu}$ using Eq.~\eqref{eq:lie}, resulting in
\begin{equation}
    \Delta g_{\mu\nu}=h_{\mu\nu}+2\nabla_{(\mu}\xi_{\nu)} \ .
\end{equation}
However, the equation for $\Delta s$ [Eq.~(21) in \cite{Friedman1978}] does not hold in our case anymore because perturbations are not adiabatic when including viscosity, as pointed out already in \cite{Hiscock1983}.
Moreover, the equation for $\Delta n$ [Eq.~(26) of said work] does not hold for arbitrary theories where baryon number is not conserved.

The equations of motion are the \textit{Eulerian} perturbation of the Einstein field equations, and the \textit{Lagrangian} perturbation of the conservation of the stress-energy tensor and the baryon number current, i.e., 
\begin{subequations}
    \begin{align}
        \delta G_{\mu\nu}&=8\pi\delta T_{\mu\nu} \ ,\\
        \Delta(\nabla_\nu T^{\mu\nu})&=0 \ ,\\
        \Delta(\nabla_\mu J^\mu)&=0 \ .
    \end{align}
\end{subequations}
The equation for conservation of the stress-energy tensor can be decomposed into components parallel and perpendicular to the four velocity, whose resulting equation is equivalent to simply taking the Lagrangian variation of Eqs.~\eqref{eq:para} and \eqref{eq:perp} respectively.

We now consider the fact that we study perturbations about equilibrium backgrounds, which include a static and spherically symmetric star, and a stationary and axisymmetric star.
For an equilibrium configuration, all the viscous fluxes are identically zero, and thus, $\mathscr E=\mathscr P=\mathscr Q^\mu=\pi^{\mu\nu}=0$ (recalling that these are background quantities).
Their perturbations, however, are not necessarily zero.
Similarly, the quantities $\sigma^{\mu\nu},\Theta$, and $\nabla_\mu\varphi$ must vanish in the equilibrium background, and also 
\begin{equation}
    \Pi^{\mu\lambda}\nabla_\lambda\log T+a^\mu=0 \ ,
\end{equation}
can be shown to be hold in the background \cite{RezzollaZanotti2013}.
Moreover, the terms inside the square brackets of Eqs.~\eqref{eq:bdnke}-\eqref{eq:bdnkq} are nothing but the parallel projection and perpendicular projection of the divergence of the perfect-fluid stress-energy tensor,
which must vanish in equilibrium.
All of this means that the terms in the constitutive relations,
Eq.~\eqref{eq:bdnk}, can be written as (the sum of) terms made up of a transport coefficient multiplying a quantity that is zero in the background.
Because of this, we assume from now on that the transport coefficient are functions only of the background functions.
We can make the same assumption for the Eckart [Eqs.~\eqref{eq:Eckart}] and the Maxwell-Cattaneo [Eqs.~\eqref{eq:mc}] constitutive relation via the same reasoning.

With this in mind and keeping only first-order terms, the first-order perturbations of Eqs.~\eqref{eq:para}-\eqref{eq:bary} are given by
\begin{widetext}
\begin{subequations}
    \begin{align}
        \Delta u^\mu\nabla_\mu\varepsilon+u^\mu\nabla_\mu\Delta\varepsilon+(\varepsilon+p)\Delta\Theta
        &=u^\mu\nabla_\mu\Delta\mathscr E-\Delta\mathscr Q_\mu a^\mu -\nabla_\mu\Delta\mathscr Q^\mu \ , \label{eq:ppara}\\
        (\Delta\varepsilon+\Delta p)a^\mu+(\varepsilon+p)\Delta a^\mu+\Delta(\Pi^{\mu\lambda}\nabla_\lambda p)&=-(\Delta\mathscr E+\Delta \mathscr P)a^\mu-\Pi^{\mu\lambda}\nabla_\lambda\Delta\mathscr P-\Pi^{\mu\lambda}\nabla_\nu(\Delta\pi^\nu_\lambda)
        -u^\nu\nabla_\nu\Delta\mathscr Q^\mu \ , \label{eq:pperp}\\
        n\Delta\Theta+\Delta u^\mu\nabla_\mu n+u^\mu\nabla_\mu\Delta n
        &=-u^\mu\nabla_\mu\Delta\mathscr N-\nabla_\mu\Delta\mathscr J^\mu \ . \label{eq:pbary}
    \end{align}
\end{subequations}
\end{widetext}
To simplify the above expressions, we must write $\Delta\varepsilon$ and $\Delta p$ in terms of $\Delta n$ and $\Delta s$. We do so using
\begin{align}
    \Delta\varepsilon&=\frac{\varepsilon+p}{n}\Delta n+nT\Delta s \ , \label{eq:devare}\\
    \Delta p&=\frac{\gamma p}{n}\Delta n+nTc_n^2\Delta s \ , \label{eq:devarp}
\end{align}
where $\gamma \equiv\partial \log p/\partial \log n\eval_s$ is known as the adiabatic index, and is related to the adiabatic speed of sound by
\begin{equation}
    \gamma p=(\varepsilon+p)c_s^2,\quad c_s^2\equiv \frac{\partial p}{\partial\varepsilon}\eval_s \ ,
\end{equation}
and the quantity $c_n^2$ is defined as
\begin{equation}
    c_n^2\equiv \frac{\partial p}{\partial\varepsilon}\eval_n \,.
\end{equation}
Although $c_n^2$ looks similar to the adiabatic speed of sound $c_s^2$, the former is not necessarily positive nor less than $1$.
Moreover, we also need expressions for $\Delta\varphi$ and $\Delta T$ in terms of $\Delta n$ and $\Delta s$ to evaluate the perturbations of the constitutive relation for the heat flux.
Both of these can be written as
\begin{equation}
    \Delta T=\frac{\partial T}{\partial n}\eval_s\Delta n+\frac{\partial T}{\partial s}\eval_n\Delta s,\quad \Delta\varphi=\frac{\partial \varphi}{\partial n}\eval_s\Delta n+\frac{\partial \varphi}{\partial s}\eval_n\Delta s \ ,
\end{equation}
where the coefficients are 
\begin{subequations}
    \begin{align}
        \frac{\partial T}{\partial n}\eval_s&=\frac1nTc_n^2,&\frac{\partial T}{\partial s}\eval_n&=\frac{1}{n}\frac{\partial^2\varepsilon}{\partial s^2}\eval_n \ ,\\
        \frac{\partial\varphi}{\partial n}\eval_s&=\frac{\varepsilon+p}{n^2T}\left[c_s^2-c_n^2\right], &\frac{\partial \varphi}{\partial s}\eval_n&=c_s^2-\frac{\varepsilon+p}{nT^2}\frac{\partial T}{\partial s}\eval_n \ ,
    \end{align}
\end{subequations}
as determined from thermodynamic relations (many useful ones are found in \cite{Hiscock1983}). The change of variables to use $\Delta n$ and $\Delta s$ is particularly helpful in that combining Eqs.~\eqref{eq:ppara} and~\eqref{eq:pbary} results in a simple equation for $\Delta s$,
\begin{align}\label{eq:entropy_m}
    nTu^\mu\nabla_\mu\Delta s=&\frac{\varepsilon+p}{n}\left[\nabla_\mu\Delta\mathscr J^\mu+u^\mu\nabla_\mu\mathscr N\right] \nonumber\\
    &-u^\mu\nabla_\mu\Delta\mathscr E-\Delta\mathscr Q_\mu a^\mu-\nabla_\mu\Delta\mathscr Q^\mu \ .
\end{align}
This last equation tells us that perturbations of out-of-equilibrium neutron stars are generally \textit{not} adiabatic (i.e. $u^\mu\nabla_\mu\Delta s=0$). 
However, out-of-equilibrium perturbations will be adiabatic in the specific case when the only out-of-equilibrium effect comes from the bulk and shear corrections [i.e. when $\mathscr P$ and $\pi_{\mu\nu}$ are the only non-zero viscous fluxes of Eq.~\eqref{eq:flux}].
This can be seen in, for example, \cite{Saketh2024}.

\section{Radial Perturbation Equations in Spherical Symmetry: BDNK, Eckart and MIS fluids}\label{sec:spher}

In this section, we consider a static, spherically symmetric background, and then derive the equations of motion for radial perturbations for all three fluid models.
Due to the many terms that appear in some of these equations, the full derivation for some of the equations is left to Appendix~\ref{app:der}.

\subsection{Background spacetime}

Consider the case of a static and spherically symmetric background.
We take the line element to be
\begin{equation}
    ds^2=-e^{-2\Phi(r)}dt^2+e^{2\Lambda(r)}dr^2+r^2d\Omega^2 \ .
\end{equation}
As was previously mentioned, this background is in equilibrium, and thus, it is governed by the perfect-fluid solution.
The perfect-fluid solution is well-known to be given by the Tolman-Oppenheimer-Volkoff (TOV) equations, which we write in the form
\begin{subequations}
    \begin{align}
        \frac{d}{dr}p&=(\varepsilon+p)\frac{d}{dr}\Phi \ ,\\
        \frac{d}{dr}\Phi&=
    \frac r2\left[-e^{2\Lambda}\left(8\pi p+\frac{1}{r^2}\right)+\frac{1}{r^2}\right] \ , \label{eq:phi}\\
    \frac{d}{dr}\Lambda&=
    \frac r2\left[e^{2\Lambda}\left(8\pi\varepsilon+\frac{1}{r^2}\right)-\frac{1}{r^2}\right] \ . \label{eq:lambda}
    \end{align}
\end{subequations}
Combining Eqs.~\eqref{eq:phi}-\eqref{eq:lambda} gives us an identity that we will use later on,
\begin{equation}\label{eq:rela}
    \frac{d}{dr}\Lambda-\frac{d}{dr}\Phi=4\pi re^{2\Lambda}(\varepsilon+p) \ .
\end{equation}
Importantly, this means that ${d\Lambda}/{dr}-{d\Phi}/{dr}\geq 0$, while also ${d\Phi}/{dr}\leq 0$.
The TOV equations can then be integrated with an equation of state, usually given as $p=p(\varepsilon)$.
In our case, where $p=p(n,s)$ and $\varepsilon=\varepsilon(n,s)$, in static and spherical symmetry, the neutron star is isentropic (i.e.~$\nabla_\mu s=0$). This implies that the equation of state is effectively one dimensional $p=p(n,s_0),\,\varepsilon=\varepsilon(n,s_0)$, which allows writing simply $p=p(\varepsilon)$.
The radius of the star $R$ is defined as the radial coordinate at which $p(r=R)=0$ and we will be dealing with equations of state such that this also implies $\varepsilon(R)=0$.
The outside of the star is given by the Schwarzschild solution.

\subsection{Perturbation Equations}

We will now derive the first-order system of the equations of motion that describes the radial perturbations of the star for the different relativistic viscous theories highlighted above, namely the Eckart [Eq.~\eqref{eq:Eckart}], the BDNK [Eq.~\eqref{eq:bdnk}], and the Maxwell-Cattaneo [Eq.~\eqref{eq:mc}] form of the IS equations.
The derivation follows closely the derivation of Chandrasekhar for a perfect fluid \cite{Chandrasekhar1964}.

We start by stating what elements of the Chandrasekhar derivation remain the same in our case.
The Lagrangian displacement once again has only a component in the $r$ direction, so that notation simplifies by taking $\xi^\mu=(0,\xi,0,0)$.
Moreover, the addition of viscosity does not change the relationship between the four velocity and the Lagrangian displacement, and thus we still have $\delta u^t=e^\Phi\delta\Phi$ and $\delta u^r=e^\Phi\partial_t\xi$.
We already knew this should be the case because, for general perturbations, Eq.~\eqref{eq:delumu} does not change.
Moreover, this means that the derivatives of $u^\mu$ of Eq.~\eqref{eq:deru} are left unchanged from the perfect-fluid case, and are given by
\begin{subequations}\label{eq:velperts}
    \begin{align}
        \Delta\Theta&=e^\Phi\left[\partial_t\delta\Lambda+(\tfrac2r+\Lambda'+\partial_r)\partial_t\xi\right] \ ,
        \\\Delta\sigma_r^r&=\frac23\left[\Delta\Theta-e^\Phi\tfrac3r\partial_t\xi\right] \ ,\\
        \Delta a_r&=e^{2\Lambda+2\Phi}\partial_t^2\xi+2\delta\Lambda\Phi'-\partial_r\delta\Phi \nonumber\\
        &\ +2\Lambda'\Phi'\xi-\xi\Phi''+\Phi'\partial_r\xi \ ,
    \end{align}
\end{subequations}
where the primes stand for radial derivatives, $\Delta a_t=\Delta a_\theta=\Delta a_\varphi=0,\Delta \sigma_\theta^\theta=\Delta\sigma_\varphi^\varphi=-\frac12\Delta\sigma_r^r$ and all other components of $\Delta\sigma_\mu^\nu$ are $0$.
Moreover, $\Delta\Theta=\delta\Theta$ and $\Delta\sigma_r^r=\delta\sigma_r^r$ up to first order in perturbations, because these quantities are zero in the background.

We now look at equations that do change from the perfect-fluid case.
First, we evaluate Eq.~\eqref{eq:pbary} for radial perturbations, leading to
\begin{align}
    \partial_t\Delta n+\frac{ne^{-\Phi}}{r^2}\partial_r(r^2e^{\Phi}\partial_t\xi)+&n\left[\partial_t\delta\Lambda+(\Lambda'-\Phi')\partial_t\xi\right] \nonumber\\
    &=-\partial_t\Delta \mathscr N-\nabla_\mu\Delta\mathscr J^\mu \ .
\end{align}
When there are no corrections to the baryon number current (that is, when $\mathscr N=0$ and $\mathscr J^\mu=0$ in the constitutive relations), the right-hand side of this equation vanishes and the left-hand side is then an overall time derivative.
Integrating in time implies that 
\begin{align}\label{eq:dn}
    \Delta n+\frac{ne^{-\Phi}}{r^2}\partial_r(r^2e^{\Phi} \xi)+&n\left[\delta\Lambda+(\Lambda'-\Phi')\xi\right]  = {\rm{const}}\,,
\end{align}
and the constant of integration can be set to zero through appropriate choices of initial conditions. 
The above equation simplifies the system of equations because $\Delta n$ will now be related algebraically to other variables, like $\xi$ and $\delta\Lambda$, and thus, it will decouple from the system of equations\footnote{The condition $\mathscr J^\mu=0$ is sufficient for $\Delta n$ to decouple from the rest of the system.}.
From here on, we assume $\Delta\mathscr N=0$ and $\Delta\mathscr J^\mu=0$ because this is the case for all the fluid models we examine in this paper.

The $tt$, $rr$, and $tr$ components of the perturbed Einstein equations now have additional terms from the stress-energy tensor, namely
\begin{subequations}
    \begin{align}
        \partial_r(re^{2\Lambda}\delta\Lambda)-4\pi re^{2\Lambda}\delta\varepsilon&=4\pi re^{2\Lambda}\delta\mathscr E \ ,\\
        \partial_r\delta\Phi-\delta\Lambda\left[2\Phi'-\frac1r\right]+4\pi re^{2\Lambda}\delta p&=-4\pi re^{2\Lambda}\left[\delta\mathscr P+\delta\pi_r^r\right] \ , \label{eq:eferr}\\
        \partial_t\delta\Lambda+(\Lambda'-\Phi')\partial_t\xi&=-4\pi re^{-\Phi}\delta\mathscr Q_r \ , \label{eq:dtlam}
    \end{align}
\end{subequations}
where we write the equations such that the right-hand side is zero for perfect fluids.
Of particular interest is the final equation [Eq.~\eqref{eq:dtlam}] because, for a perfect fluid, the equation is a total time derivative that can be easily integrated to find that $\delta\Lambda$ decouples from the rest of the system.
This is equivalent to what happens with $\Delta n$ when $\mathscr J^\mu=0$, but we do not consider the case $\mathscr Q^\mu=0$ here, since the BDNK relations Eq.~\eqref{eq:bdnk} always have a nonzero heat flux.

From here on, we treat each of the different fluid models separately to obtain the perturbed equations of motion.
In the end, we will rewrite each theory as a system of first-order-in-time equations that is simpler to analyze.
Going forward, we introduce the variable 
\begin{equation}
    \Xi\equiv r^2e^\Phi\xi\, ,
\end{equation} 
since it is in this variable that the Chandrasekhar eigenvalue problem is most clearly a Sturm-Liouville problem \cite{MTW,Caballero2024}.
Explicitly, the Chandrasekhar problem is given by
\begin{equation}\label{eq:chandra}
    W(r)\partial_t\dot\Xi=\partial_r(P(r)\partial_r\Xi)-Q(r)\Xi \ ,
\end{equation}
where the overhead dot stands for a time derivative, while the functions $W(r), P(r)$, and $Q(r)$ are given by
\begin{subequations}\label{eq:wpq}
    \begin{align}
        W(r)&\equiv e^{3\Lambda-\Phi}\frac{(\varepsilon+p)}{r^2}\label{eq:bW} \ ,\\
        P(r)&\equiv e^{\Lambda-3\Phi}\frac{\gamma \, p}{r^2}\, \label{eq:bP}  \ ,\\
        Q(r)&\equiv-e^{\Lambda-3\Phi}\frac{(\varepsilon+p)}{r^2}\left[\Phi''-\frac2r\Phi'-(\Phi')^2\right] \nonumber   \\
        &\quad+4\pi e^{3\Lambda-3\Phi}\frac{(\varepsilon+p)^2}{r}\left(\Phi'-\frac1r\right) . \label{eq:Qb}
    \end{align}
\end{subequations}
Notice that $\Phi$, $\Lambda$, $\varepsilon$ and $p$ are known functions from the background evolution equations, but we choose to not use the latter here to simplify the above expressions. 
The general solution to Eq.~\eqref{eq:chandra} is
\begin{equation}\label{eq:eigenexpand}
    \Xi(t,r)=\sum_je^{i\omega_jt}\phi_j(r)
\end{equation}
with $\omega_j^2$ and $\phi_j(r)$ the eigenvalues\footnote{Properly, the eigenvalues of the problem are $\omega_j^2$ so that we refer to $\omega_j$ as the eigenfrequencies.} and eigenvectors of Eq.~\eqref{eq:chandra} respectively, under the transformation $\partial_t^2\Xi(t,r)\to-\omega_j^2\phi_j(r)$ and $\Xi(t,r)\to\phi_j(r)$.
The eigenfrequencies $\omega_j$ can be either real or imaginary, and a configuration is stable if and only if all the eigenfrequencies are real.
Moreover, if there is a positive eigenvalue $\omega_0^2$ that it is smaller than all other eigenvalues, then all other eigenfrequencies are real and the configuration is stable.

In the following sections, we will use 
\begin{align}
\dot\Xi &\equiv\partial_t\Xi\,,
\\
\Psi&\equiv \delta\Lambda+(\Lambda'-\Phi')\xi\,
\end{align}
as variables of motion, such that Eq.~\eqref{eq:dtlam} can then be written as\footnote{We have that $\Delta\mathscr Q_r=\delta\mathscr Q_r$ from the fact that $\mathscr Q_\mu=0$ in the background.}
\begin{equation}\label{eq:psi}
    \partial_t\Psi=-4\pi re^{-\Phi}\Delta\mathscr Q_r \ .
\end{equation}
In the following, we can assume that all the viscous coefficients are simply functions of $r$. This is because the transport coefficients are functions of the background variables, and the background is a function of $r$ only. Thus, we can write $\zeta=\zeta(n(r),s(r))$, and, going forward, we treat all transport coefficients as functions of $r$ only.

\subsubsection{Eckart Fluid}\label{sssec:eckart}

We now derive the perturbation equations for radial oscillations for the Eckart constitutive relations in Eq.~\eqref{eq:Eckart}.
We will show that the resulting equations can be written as a first-order system in $4$ variables of motion: $(\Xi,\dot\Xi,\Delta s,\Psi)$.
To obtain the new equation of motion for $\Delta s$ we start by evaluating Eq.~\eqref{eq:entropy_m} for radial perturbations with the Eckart constitutive relations.
The only nonzero terms come from those with $\Delta\mathscr Q_\mu$ in Eq.~\eqref{eq:entropy_m} and result in
\allowdisplaybreaks[4]
\begin{equation}\label{eq:dseck}
    \partial_t\Delta s=-\frac{1}{nT}\frac{e^{-\Lambda+\Phi}}{r^2}\partial_r\left(e^{-\Lambda-2\Phi}r^2\Delta\mathscr Q_r\right) \ .
\end{equation}
With Eqs.~\eqref{eq:psi} and \eqref{eq:dseck}, and recalling that $\partial_t\Xi=\dot\Xi$,
the only remaining equation one needs to derive is that for $\partial_t\dot\Xi$. 
We obtain this equation from Eq.~\eqref{eq:pperp} and find that it is the equivalent of the Chandrasekhar problem of Eq.~\eqref{eq:chandra}.
The derivation of the equation is given in Appendix~\ref{app:der}.
The result is
\begin{widetext}
\begin{align}\label{eq:bigg2}
    W(r)\partial_t\dot\Xi=&\partial_r(P(r)\partial_r\Xi)-Q(r)\Xi \nonumber\\
    &+\partial_r\left(e^{\Lambda-2\Phi}\gamma p\Psi\right)+(\varepsilon+p)e^{\Lambda-2\Phi}\Psi\left[\Phi'-\frac1r\right] \nonumber\\
    &-\partial_r\left(e^{\Lambda-2\Phi}c_n^2nT\Delta s\right)+\Phi'e^{\Lambda-2\Phi}nT\Delta s \nonumber\\
    &-e^{\Lambda-\Phi}\partial_t\Delta\mathscr Q_r+\partial_r\left[(\zeta+\tfrac43\eta)e^{\Lambda-2\Phi}\left(\frac{\partial_r\dot\Xi}{r^2}-4\pi r\Delta\mathscr Q_r\right)\right] \nonumber\\
    &-\frac{4\eta}{r}e^{\Lambda-2\Phi}\left[(\Lambda'-2\Phi')\frac{\dot\Xi}{r^2}+4\pi r \Delta\mathscr Q_r\right] -e^{\Lambda-2\Phi}\frac{4\partial_r\eta}{r}\frac{\dot\Xi}{r^2}\ ,
\end{align}
\end{widetext}
where $W,P,Q$ are the same as in Eq.~\eqref{eq:wpq} and $\Delta\mathscr Q_r$ is an expression that depends only on $\Xi$ and $\Delta s$, and it is given by
\begin{align}\label{eq:meckq}
    \Delta\mathscr Q_r=\kappa T\frac{\varepsilon+p}{n}\partial_r\bigg[&-\frac{\varepsilon+p}{nT}(c_s^2-c_n^2)\left(\frac{e^{-\Phi}}{r^2}\partial_r\Xi+\Psi\right) \nonumber\\
    &+\left(c_s^2-\frac{\varepsilon+p}{n}\frac{1}{T^2}\frac{\partial T}{\partial s}\eval_n\right)\Delta s\bigg] \ .
\end{align}
Notice that although the term $\partial_t\Delta\mathscr Q_r$ introduces terms of the form $\partial_t\Psi$ and $\partial_t\Delta s$, we can eliminate such terms using Eqs.~\eqref{eq:psi} and \eqref{eq:dseck}, which results in a right-hand side with no time derivatives.
The quantities in the second and third lines of Eq.~\eqref{eq:bigg2} are not present in the perfect-fluid case, even though they are not multiplied by any viscous terms.
This is because for a perfect fluid both $\partial_t\Psi=0,$ and $\partial_t\Delta s=0$, so that, for appropriately-chosen initial conditions, one has $\Delta s=0$ and $\Delta\Psi=0$.
Equations \eqref{eq:bigg2} together with Eqs.~\eqref{eq:psi}, \eqref{eq:dseck} and $\partial_t\Xi=\dot\Xi$, form a first-order system in four variables $(\Xi,\dot\Xi,\Psi,\Delta s)$
that govern the radial perturbations for a non-rotating neutron star for an Eckart fluid.

\subsubsection{BDNK Fluid}

For the BDNK equations, the simplest way to obtain a first-order system is to make $\Delta \mathscr E$ and $\Delta \mathscr Q_r$ variables of motion,
while $\Delta\mathscr P$ will be related to the other quantities by $\Delta\mathscr P=-\zeta\Delta\Theta+{\tau_\mathscr P}/{\tau_\mathscr E}\Delta\mathscr E$.
The reason for treating $\Delta\mathscr E$ and $\Delta\mathscr Q_r$ as variables of motion is that this allows us to write the perturbation of Eq.~\eqref{eq:bdnke} as
\begin{equation}\label{eq:ds}
    \tau_{\mathscr E}nTe^\Phi\partial_t\Delta s=\Delta \mathscr E \ ,
\end{equation}
which is a first-order equation for $\Delta s$.
At the same time, Eq.~\eqref{eq:entropy_m} for the radial perturbations can be written as
\begin{equation}\label{eq:de}
    \partial_t\Delta\mathscr E=-\frac{e^{-\Phi}}{\tau_{\mathscr E}}\Delta\mathscr E-\frac{e^{-\Lambda+\Phi}}{r^2}\partial_r\left(e^{-\Lambda-2\Phi}r^2\Delta\mathscr Q_r\right) \ .
\end{equation}

The last equation of motion that governs the perturbations is Eq.~\eqref{eq:pperp}, but it has two unknowns $u^\nu \nabla_\nu\Delta\mathscr Q_\mu$ and $u^\nu\nabla_\nu\Delta u^\mu$.
For this reason, we must also use Eq.~\eqref{eq:bdnkq} to find individual equations for $u^\nu\nabla_\nu\Delta\mathscr Q_\mu$ and $u^\nu\nabla_\nu \Delta u^\mu$.
We know that the term in square brackets multiplying $\tau_{\mathscr Q}$ in Eq.~\eqref{eq:bdnkq} is the left-hand side of Eq.~\eqref{eq:pperp}.
Then, solving for $\Delta[(\varepsilon+p)a_\mu+\Pi_{\mu}^{\lambda}\nabla_\lambda p]$ in Eq.~\eqref{eq:bdnkq} and substituting into Eq.~\eqref{eq:pperp}, we obtain the following first-order equation for $\Delta\mathscr Q^\mu$:
\begin{align}
    &u^\nu\nabla_\nu\Delta\mathscr Q^\mu+\frac{1}{\tau_{\mathscr Q}}\left[\Delta\mathscr Q^\mu-\kappa T\frac{\varepsilon+p}{n}\Pi^{\mu\lambda}\nabla_\lambda\Delta\varphi\right] \nonumber\\
    =-&(\Delta\mathscr E+\Delta\mathscr P)a^\mu-\Pi^{\mu\lambda}\nabla_\lambda\Delta\mathscr P-\Pi^{\mu\lambda}\nabla_\nu(\Delta\pi^\nu_\lambda) \ ,
\end{align}
which when evaluated for radial perturbations (where $\mu=r$ is the only nonzero component of the heat flux) yields
\begin{widetext}
\begin{align}\label{eq:dq}
    \partial_t\Delta\mathscr Q_r=&-\frac{e^{-\Phi}}{\tau_{\mathscr Q}}\left[\Delta\mathscr Q_r-\kappa T\frac{\varepsilon+p}{n}\partial_r\delta\varphi\right]+e^{-\Phi}\Phi'\Delta\mathscr E-\partial_r\left(e^{-\Phi}\frac{\tau_{\mathscr P}}{\tau_{\mathscr E}}\Delta\mathscr E\right) \nonumber\\
    &+\partial_r\left[(\zeta+\tfrac43\eta)\left(\frac{e^{-\Phi}}{r^2}\partial_r\dot\Xi-4\pi re^{-\Phi}\Delta\mathscr Q_r\right)\right]+\frac{4\eta}{r}\left[\frac{e^{-\Phi}}{r^2}\Phi'\dot\Xi-4\pi re^{-\Phi}\Delta\mathscr Q_r\right] +\frac{4\partial_r\eta}{r}\frac{e^{-\Phi}\dot\Xi}{r^2}\ .
\end{align}
\end{widetext}
Finally, our remaining equation of motion is found by solving for $\partial_t\dot\Xi$ through the $r$ component of Eq.~\eqref{eq:pperp}, which results in Eq.~\eqref{eq:biggo}, and then substituting Eq.~\eqref{eq:dq} for $\partial_t\Delta\mathscr Q_r$.
After canceling some terms and simplifying, this results in
\begin{widetext}
    \begin{align}\label{eq:biggg}
    W(r)\partial_t\dot\Xi=&\partial_r(P(r)\partial_r\Xi)-Q(r)\Xi \nonumber\\
        &+\partial_r\left(e^{\Lambda-2\Phi}\gamma p\Psi\right)+(\varepsilon+p)e^{\Lambda-2\Phi}\Psi\left[\Phi'-\frac1r\right] \nonumber\\
        &-\partial_r\left(e^{\Lambda-2\Phi}c_n^2nT\Delta s\right)+\Phi'e^{\Lambda-2\Phi}nT\Delta s \nonumber\\
        &-\frac{e^{\Lambda-2\Phi}}{\tau_{\mathscr Q}}\left[\Delta\mathscr Q_r-\kappa T\frac{\varepsilon+p}{n}\partial_r\delta\varphi\right]-16\pi^2r^2e^{3\Lambda-2\Phi}(\varepsilon+p)(\zeta+\tfrac43\eta)\Delta\mathscr Q_r \nonumber\\
        &-4\pi r(\varepsilon+p)e^{3\Lambda-2\Phi}\left[\frac{\tau_{\mathscr P}}{\tau_{\mathscr E}}\Delta\mathscr E+\frac{4\eta}{r^3}\dot\Xi-\frac{(\zeta+\tfrac43\eta)}{r^2}\partial_r\dot\Xi\right] \ ,
    \end{align}
\end{widetext}
where $W,P$, and $Q$ are the same as in Eq.~\eqref{eq:wpq}.

The system of equations Eqs.~\eqref{eq:psi}, \eqref{eq:ds}, \eqref{eq:de}, \eqref{eq:dq}, \eqref{eq:biggg}, together with $\partial_t\Xi=\dot\Xi$,
is a first-order system of $6$ variables $(\Xi,\dot\Xi,\Delta s,\Psi,\Delta\mathscr E,\Delta\mathscr Q_r)$.
Notice that the first three lines of Eq.~\eqref{eq:biggg} are the same as those of Eq.~\eqref{eq:bigg2}.
Moreover, if we take the limit $(\tau_\mathscr E,\tau_\mathscr P,\tau_\mathscr Q)\to 0$\footnote{This requires using Eqs.~\eqref{eq:de} and \eqref{eq:dq} to avoid dividing by zero.}, Eq.~\eqref{eq:biggg} becomes Eq.~\eqref{eq:bigg2}, while Eq.~\eqref{eq:ds} becomes Eq.~\eqref{eq:dseck},
i.e.~if we remove the relaxation times, the system becomes an Eckart fluid, as expected.

\subsubsection{Maxwell-Cattaneo Fluid}

Lastly, we derive the equations for radial perturbations for the Maxwell-Cattaneo equations in the form of a first-order system, so that we can compare with the BDNK case.
The quantities $\Delta\mathscr P,\Delta\pi_\mu^\nu$, and $\Delta\mathscr Q_\mu$ are now variables of motion,
but only $\Delta \mathscr P,\Delta \mathscr Q_r,\Delta\pi_r^r,\Delta\pi_\theta^\theta$, and $\Delta\pi_\varphi^\varphi$ are nonzero.
Moreover, $\Delta\pi_r^r+\Delta\pi_\theta^\theta+\Delta\pi_\varphi^\varphi=0$, so that instead of writing equations for $\Delta\pi_\theta^\theta$ and $\Delta\pi_\varphi^\varphi$, we only need an equation for $\Delta\pi_\theta^\theta-\Delta\pi_\varphi^\varphi$, which is in fact zero from the Einstein equations\footnote{Specifically, as a consequence of spherical symmetry, $G_\theta^\theta-G_\varphi^\varphi$ is zero. Then, from the Einstein equations, $\Delta \pi_\theta^\theta-\Delta \pi^\varphi_\varphi=0$.}.
Therefore, we obtain three independent equations for $\Delta\mathscr P,\Delta\mathscr Q_r$, and $\Delta\pi_r^r$.
The equations of motion for these variables are all relaxation equations with a source, which we write explicitly as
\begin{subequations}\label{eq:misrel}
\begin{align}
    \tau_0 e^\Phi\partial_t\Delta\mathscr P+\Delta\mathscr P&=-\zeta e^\Phi\left[-4\pi re^{-\Phi}\Delta\mathscr Q_r+\frac{e^{-\Phi}}{r^2}\partial_r\dot\Xi\right] \ , \label{eq:misp}\\
    \tau_1e^\Phi\partial_t\Delta\mathscr Q_r+\Delta\mathscr Q_r&=\kappa T\frac{\varepsilon+p}{n}\partial_r\delta\varphi \ , \label{eq:misq}\\
    \tau_2e^\Phi\partial_t\Delta\pi_r^r+\Delta\pi_r^r&=-\frac43\eta e^\Phi\Big[-4\pi re^{-\Phi}\Delta\mathscr Q_r \nonumber\\
    &\quad\quad\quad\quad\quad+\frac{e^{-\Phi}}{r^2}\partial_r\dot\Xi-\frac{3}{r}\frac{e^{-\Phi}}{r^2}\dot\Xi\Big]  \ .\label{eq:mispi}
\end{align}
\end{subequations}
Equations \eqref{eq:psi} and \eqref{eq:dseck} are once again necessary equations of motion.
The final equation of motion we once again derive in App.~\ref{app:der} and is given by
\begin{widetext}
\begin{align}\label{eq:bigg3}
    W(r)\partial_t\dot\Xi=&\partial_r(P(r)\partial_r\Xi)-Q(r)\Xi \nonumber\\
    &+\partial_r\left(e^{\Lambda-2\Phi}\gamma p\Psi\right)+(\varepsilon+p)e^{\Lambda-2\Phi}\Psi\left[\Phi'-\frac1r\right] \nonumber\\
    &-\partial_r\left(e^{\Lambda-2\Phi}c_n^2nT\Delta s\right)+\Phi' e^{\Lambda-2\Phi}nT \Delta s \nonumber\\
    &+\frac{e^{\Lambda-2\Phi}}{\tau_1}\left[\Delta\mathscr Q_r-\kappa T\frac{\varepsilon+p}{n}\partial_r\delta\varphi\right]-\partial_r\left[e^{\Lambda-2\Phi}(\Delta\mathscr P+\Delta\pi_r^r)\right]-\frac3re^{\Lambda-2\Phi}\Delta\pi_r^r \ .
\end{align}
\end{widetext}

The resulting system is quite similar to that produced by a BDNK fluid in that we have an equation that is a modification of the Chandrasekhar equation [Eq.~\eqref{eq:biggg} for a BDNK fluid and Eq.~\eqref{eq:bigg3} for a MIS fluid], two auxiliary equations for $\Psi$ [Eq.~\eqref{eq:psi} for both] and $\Delta s$ [Eq.~\eqref{eq:ds} for a BDNK fluid and Eq.~\eqref{eq:dseck} for a MIS fluid], and a set of equations with a relaxation term [Eqs.~\eqref{eq:de} and \eqref{eq:dq} for a BDNK fluid and Eqs.~\eqref{eq:misrel} for a MIS fluid].

\section{Small viscosity for radial problem}\label{sec:small}

In the previous section, we derived first-order systems that describe the equations of motion of the perturbations for each of the Eckart, BDNK and MIS equations.
Since all these systems have no explicit time dependence, a harmonic decomposition should lead to eigenvalue problems for the radial modes of oscillation.
Therefore, the stability of the radial mode can be determined by calculating the eigenvalues of this problem.
However, the resulting systems of equations are quite complicated. For example, the operator for the Eckart fluid that describes the time evolution will be a $4 \times 4$ matrix of differential and multiplicative operators.
Moreover, the resulting matrix operator is not expected to be self-adjoint, making it difficult to establish mode completeness.
For this reason, we limit ourselves to studying the stability of these fluids in the case of small viscosity, where we can rigorously establish linear radial stability by calculating corrections to the eigenvalues of the perfect-fluid problem.

We will perform the small viscosity approximation by rescaling the viscous coefficients by a small parameter $\alpha$.
Assuming that the rescaling parameter is small is equivalent to assuming that viscous effects are smaller than equilibrium effects by a factor of the small parameter.
For example, assuming the parameter is small, we ensure that $\mathscr P/p\sim \alpha$, and similar scalings for other out-of-equilibrium effects, when compared to equilibrium effects.
Alternatively, this can be understood as an assumption that the Knudsen number (i.e.~the ratio between the microscopic and the macroscopic scales of the system) is small and that $\alpha$ is proportional to the Knudsen number.

\subsection{Eckart Fluid}\label{ssec:smalleck}

We now carry out a small viscosity expansion for the equations of motion for the Eckart problem (i.e.~those given in Sec.~\ref{sssec:eckart}).
We rescale all viscous coefficients via $(\zeta,\eta,\kappa)\to (\alpha\zeta,\alpha\eta,\alpha\kappa)$ for $\alpha \ll 1$,
and expand each variable of motion in a series in $\alpha$, i.e.~we write
\begin{equation}\label{eq:smallexp}
    \Xi(t,r)=\Xi^{(0)}(t,r)+\alpha\Xi^{(1)}(t,r)+\dots \ ,
\end{equation}
for each variable $\Xi,\dot\Xi,\Psi$, and $\Delta s$. At zeroth order, Eqs.~\eqref{eq:psi} and \eqref{eq:dseck} result in $\partial_t\Psi^{(0)}=0$ and $\partial_t\Delta s^{(0)}=0$, following from the fact that $\Delta\mathscr Q_r$ is at least ${\cal{O}}(\alpha)$ from Eq.~\eqref{eq:meckq}.
By choosing appropriate initial conditions, like in the Chandrasekhar problem,
we can set $\Psi^{(0)}=\Delta s^{(0)}=0$.
This leads to Eq.~\eqref{eq:bigg2}, which reduces to simply the Chandrasekhar problem of Eq.~\eqref{eq:chandra}, as expected. The solution to this problem was already presented in Eq.~\eqref{eq:eigenexpand}, and we will label the eigenfrequencies and eigenvectors by $\omega_j^{(0)}$ and $\phi_j^{(0)}$ respectively.

Moving on to first order in $\alpha$, the expressions for $\partial_t\Psi^{(1)}$ and $\partial_t\Delta s^{(1)}$ will be proportional to $\Delta\mathscr Q_r^{(1)}$, which can be obtained by just evaluating Eq.~\eqref{eq:meckq} at first order, to find
\begin{equation}
    \Delta\mathscr Q_r^{(1)}=-\kappa T\frac{\varepsilon+p}{n}\partial_r\left[\frac{\varepsilon+p}{nT}(c_s^2-c_n^2)\frac{e^{-\Phi}}{r^2}\partial_r\Xi^{(0)}\right] \ ,
\end{equation}
where we used that $\Psi^{(0)}=\Delta s^{(0)}=0$.
Therefore, Eq.~\eqref{eq:psi} at first order is given by
\begin{equation}
    \partial_t\Psi^{(1)}=4\pi re^{-\Phi}\kappa T\frac{\varepsilon+p}{n}\partial_r\left[\frac{\varepsilon+p}{nT}(c_s^2-c_n^2)\frac{e^{-\Phi}}{r^2}\partial_r\Xi^{(0)}\right] \ .
\end{equation}
We can then expand the $\Xi^{(0)}$ into the Chandrasekhar eigenvectors according to Eq.~\eqref{eq:eigenexpand} and integrate each term of the sum independently.
We then find that $\Psi^{(1)}$ 
is given by the following expansion in the eigenvectors $\phi_j^{(0)}$ of the Chandrasekhar problem:
\begin{align}\label{eq:psiexp}
    \Psi^{(1)}=-i\sum_j \frac{e^{i\omega_j^{(0)}t}}{\omega_j^{(0)}}&4\pi re^{-\Phi}\kappa T\frac{\varepsilon+p}{n} \nonumber\\
    &\times\partial_r\left[\frac{\varepsilon+p}{nT}(c_s^2-c_n^2)\frac{e^{-\Phi}}{r^2}\partial_r\phi_j^{(0)}\right] \ ,
\end{align}
and a similar reasoning leads to the following expression for $\Delta s^{(1)}$:
\begin{align}\label{eq:sexp}
    \Delta s^{(1)}=i\sum_j\frac{e^{i\omega_j^{(0)}t}}{\omega_j^{(0)}}&\frac{1}{nT}\frac{e^{-\Lambda+\Phi}}{r^2} \nonumber\\
    \times&\partial_r\Bigg\{e^{-\Lambda-2\Phi}r^2 \kappa T\frac{\varepsilon+p}{n} \nonumber\\
    &\quad\times\partial_r\left[\frac{\varepsilon+p}{nT}(c_s^2-c_n^2)\frac{e^{-\Phi}}{r^2}\partial_r\phi_j^{(0)}\right]\Bigg\} \ .
\end{align}

We now move on to evaluating the equation of motion [Eq.~\eqref{eq:bigg2}] at first order in the expansion parameter.
We write explicitly the expression at first order as
\begin{widetext}
\begin{align}\label{eq:exp1}
    W(r)\partial_t\dot\Xi^{(1)}-\partial_r\left(P(r)\partial_r\Xi^{(1)}\right)+Q(r)\Xi^{(1)}=&+\partial_r\left(e^{\Lambda-2\Phi}\gamma p\Psi^{(1)}\right)+(\varepsilon+p)e^{\Lambda-2\Phi}\Psi^{(1)}\left[\Phi'-\frac1r\right] \nonumber\\
    &-\partial_r\left(e^{\Lambda-2\Phi}c_n^2nT\Delta s^{(1)}\right)+\Phi'e^{\Lambda-2\Phi}nT\Delta s^{(1)} \nonumber\\
    &+e^{\Lambda-\Phi}\kappa T\frac{\varepsilon+p}{n}\partial_r\left[\frac{\varepsilon+p}{nT}(c_s^2-c_n^2)\frac{e^{-\Phi}\partial_r\dot\Xi^{(0)}}{r^2}\right] \nonumber\\
    &+\partial_r\left[(\zeta+\tfrac43\eta)e^{\Lambda-2\Phi}\frac{\partial_r\dot\Xi^{(0)}}{r^2}\right]-\frac{4\eta}{r}e^{\Lambda-2\Phi}(\Lambda'-2\Phi')\frac{\dot\Xi^{(0)}}{r^2} \nonumber\\
    &-\frac{4\partial_r\eta}{r}e^{\Lambda-2\Phi}\frac{\dot\Xi^{(0)}}{r^2}\ .
\end{align}
\end{widetext}
Since $\Psi^{(1)},\Delta s^{(1)}$, and $\dot\Xi^{(0)}$ can be expanded in terms of the eigenvectors of the perfect-fluid problem,
this equation is the same as the perfect-fluid problem of Eq.~\eqref{eq:chandra}, but with a source that depends only on the solution at zeroth order, as is expected from perturbation theory.
We can then expand $\Xi^{(1)}$ in a mode expansion to find corrections to the eigenfrequencies and eigenvectors at first order, $\omega_j^{(1)}$ and $\phi_j^{(1)}$ respectively.
Seeing that the first two lines on the right-hand side of Eq.~\eqref{eq:exp1} go as $-{i} \phi_j^{(0)}/{\omega_j^{(0)}}$ [by virtue of Eqs.~\eqref{eq:psiexp} and \eqref{eq:sexp}],
while the last two lines go as $i\omega_j^{(0)}\phi_j^{(0)}$,
it is convenient to treat these terms separately.

Let us now solve this equation using techniques from quantum mechanics. 
We first introduce the inner product $\braket{\cdot}$ via
\begin{equation}
    \braket{\phi}{\psi}=\int_0^R dr\, W(r)\bar\phi\psi \ .
\end{equation}
Then, by taking the inner product of the mode expansion of Eq.~\eqref{eq:exp1} with $\phi_j^{(0)}$, we find that the correction to the eigenfrequencies is given by
\begin{equation}\label{eq:evcorr}
    \omega_j^{(1)}=\frac i2\frac{\braket{\phi_j^{(0)}}{\mathbb F\phi_j^{(0)}}}{\braket{\phi_j^{(0)}}}+\frac{i}{2}\frac{1}{\left(\omega_j^{(0)}\right)^2}\frac{\braket{\phi_j^{(0)}}{\mathbb G\phi_j^{(0)}}}{\braket{\phi_j^{(0)}}} \ ,
\end{equation}
while the correction to the eigenvector $\phi_j^{(1)}$ is given by
\begin{align}
    \phi_j^{(1)}=-i\sum_{k\ne j}&\frac{\omega_j^{(0)}}{\left(\omega_k^{(0)}\right)^2-\left(\omega_j^{(0)}\right)^2}\left[\frac{\braket{\phi_k^{(0)}}{\mathbb F\phi_j^{(0)}}}{\braket{\phi_k^{(0)}}} \right. \nonumber\\
    &\quad\left.+\frac{1}{\left(\omega_j^{(0)}\right)^2}\frac{\braket{\phi_k^{(0)}}{{\mathbb G\phi_j^{(0)}}}}{\braket{\phi_k^{(0)}}}\right]\phi_k^{(0)} \ .
\end{align}
The operators $\mathbb F$ and $\mathbb G$ are given by the (negative) of the last three lines and the first two lines of the right-hand side of Eq.~\eqref{eq:exp1} respectively. Explicitly, the operators $\mathbb F$ and $\mathbb G$ are
\begin{align}
    W(r)\,\mathbb F\phi=& -\partial_r\left[(\zeta+\tfrac43\eta)e^{\Lambda-2\Phi}\frac{\partial_r\phi}{r^2}\right] \nonumber\\
    &+\frac{4\eta}{r}e^{\Lambda-2\Phi}(\Lambda'-2\Phi')\frac{\phi}{r^2}-\frac{4\partial_r\eta}{r}e^{\Lambda-2\Phi}\frac{\phi}{r^2}\nonumber\\
    &-e^{\Lambda-\Phi}\kappa T\frac{\varepsilon+p}{n}\partial_r\left[\frac{\varepsilon+p}{nT}(c_s^2-c_n^2)\frac{e^{-\Phi}\partial_r\phi}{r^2}\right] \ , \label{eq:fop}\\
    W(r)\,\mathbb G\phi=&-\partial_r\left(e^{\Lambda-3\Phi}4\pi r\gamma p Y[\phi]\right) \nonumber\\
    &-(\varepsilon+p)e^{\Lambda-3\Phi}4\pi \left(r\Phi'-1\right)Y[\phi] \nonumber\\
    &-\partial_r\left[\frac{e^{-\Phi}c_n^2}{r^2}\partial_r\left(e^{-\Lambda-2\Phi}r^2Y[\phi]\right)\right] \nonumber\\
    &+\Phi'\frac{e^{-\Phi}}{r^2}\partial_r\left(e^{-\Lambda-2\Phi}r^2Y[\phi]\right) \ ,
\end{align}
where
\begin{equation}
    Y[\phi]=-\kappa T\frac{\varepsilon+p}{n}\partial_r\left[\frac{\varepsilon+p}{nT}(c_s^2-c_n^2)\frac{e^{-\Phi}}{r^2}\partial_r\phi\right] \ .
\end{equation}

We can now use Eq.~\eqref{eq:evcorr} to determine the radial stability of a neutron star to viscosity for small viscosity.
We assume that $\left(\omega_j^{(0)}\right)^2>0$ for all $j$, meaning that the star is radially stable before including viscosity.
To determine that a configuration is stable to viscosity, the right-hand side of Eq.~\eqref{eq:evcorr} must have a positive imaginary part, so that $\Re(i\omega_j^{(1)})<0$ and the perturbation leads to decay in the mode.
In the case of a zero imaginary part in the correction to the eigenvalues, it is not possible to state whether the configuration is stable or not, and it would require calculating the correction to the eigenvalue to next order in small viscosity to determine stability.
However, we will find that generally --i.e. as long as not all coefficients are zero or some terms perfectly cancel out-- there will be a nonzero imaginary part and stop our analysis at first order.
In what follows, we will study these stability conditions by separating the case when the heat conductivity is zero from the case when this is not true.

\subsubsection{Zero heat conductivity}\label{sssec:bulkshear}

Let us first consider the case where $\kappa=0$, i.e.~there is no heat conductivity.
In such cases, $\mathbb G\phi_j^{(0)}=0$ while
\begin{widetext}
    \begin{align}\label{eq:stablvis}
    \braket{\phi_j^{(0)}}{\mathbb F\phi_j^{(0)}}=\int_0^Rdr\, &(\zeta+\tfrac43\eta)\frac{e^{\Lambda-2\Phi}}{r^2}\left|\partial_r\phi_j^{(0)}\right|^2 +\frac{4e^{\Lambda-2\Phi}}{r^3}\left[\eta(\Lambda'-2\Phi')+\partial_r\eta\right]\left|\phi_j^{(0)}\right|^2 \nonumber\\
    -\bar\phi_j^{(0)}&(\zeta+\tfrac43\eta)e^{\Lambda-2\Phi}\frac{\partial_r\phi_j^{(0)}}{r^2}\eval_0^R \ ,
\end{align}
\end{widetext}
where the last term is a boundary term that comes from integrating by parts.
To obtain a vanishing boundary term, we must impose the condition
\begin{equation}
    \lim_{r\to R}\frac{(\zeta+\tfrac43\eta)}{r^2}\partial_r\phi_j^{(0)}=0 \ .
\end{equation}
Recall that the boundary condition at $R$ is that $({\gamma p})/{r^2}\partial_r\phi_j^{(0)}\to 0$.
Therefore, we must have that $(\zeta,\eta)\to 0$ at least as fast as $\gamma p\to 0$ at the boundary of the star.
Once we have set the boundary term to zero, the fact that $(\zeta,\eta)\geq 0$ implies that the first term in the first line is positive definite, so we can only have an instability if the second term is negative.
In fact, one can show that with this boundary condition, Eq.~\eqref{eq:stablvis} is positive definite since the right-hand side can be written as
\begin{align}\label{eq:perfectsquare}
    \braket{\phi_j^{(0)}}{\mathbb F\phi_j^{(0)}}=\int_0^Rdr\, &\zeta\frac{e^{\Lambda-2\Phi}}{r^2}\left|\partial_r\phi_j^{(0)}\right|^2 \nonumber\\
    &+\frac43\eta\frac{e^{\Lambda-2\Phi}}{r^2}\left|\partial_r\phi_j^{(0)}-\frac3r\phi_j^{(0)}\right|^2 \ .
\end{align}
To see that the two expressions are equal, one must expand the square in the second line of Eq.~\eqref{eq:perfectsquare} and use integration by parts to transform the mixed term with $\eta\bar\phi_j^{(0)}\partial_r\phi^{(0)}$ (and its complex conjugate) to a term of the form $\eta|\phi_j^{(0)}|^2$ while setting the boundary term to $0$.
This then means that $\mathbb F$ is a positive operator, and thus it can only lead to decay behavior and a neutron star is always radially mode stable to bulk and shear viscosity.

These results are in fact true even for large viscosity as long as there is no heat conductivity.
This is based on energy arguments, which are commonly used to study the stability of neutron stars \cite{Friedman1978,Lindblom1983,Prabhu_2016,Kandrup1982}.
When $\kappa=0$, then $\partial_t\Delta s=0$ and $\partial_t\Psi=0$, with well-chosen initial conditions, and so $\Delta s=0$ and $\Psi=0$.
Then, the problem reduces to a single equation of motion, Eq.~\eqref{eq:bigg2}, which takes the form $W\ddot\Xi+(W\mathbb C)\Xi+(W\mathbb F)\dot\Xi=0$,
where $W\mathbb F$ is just the first two lines of Eq.~\eqref{eq:fop}, and $W\mathbb C$ is the Sturm-Liouville operator of the perfect-fluid problem, given by the right-hand side of Eq.~\eqref{eq:chandra}.
The energy of a perturbation is defined as $E=\braket{\dot\Xi}+\braket{\Xi}{\mathbb C\Xi}$.
Using that $\mathbb C$ and $\mathbb F$ are self-adjoint, and using the equation of motion, we obtain that ${dE}/{dt}=-2\braket{\dot\Xi}{\mathbb F\dot\Xi}$.
Then, the energy is always decreasing as a consequence that $\mathbb F$ is a positive operator.
It follows that the solution is stable if the initial energy is positive for all initial data, which is the same condition for stability as the perfect-fluid problem.
Therefore, a configuration with only bulk and shear viscosity (i.e.~without heat conductivity or additional degrees of freedom such as those introduced for BDNK and MIS fluids) is radially mode stable if and only if both $\mathbb C$ and $\mathbb F$ have only positive eigenvalues, regardless of the amount of viscosity in the system. 
We note this result is not quite new in that the expression in Eq.~\eqref{eq:perfectsquare} is the same as evaluating the time derivative of the energy functional given in \cite{Kandrup1982,Lindblom1983} for radial perturbations.
However, this is the first time that it has been explicitly written in terms of mode solutions with the intention of calculating the eigenfrequencies.
Moreover, our result applies to any radial perturbations, while the results of \cite{Kandrup1982,Lindblom1983} were derived in the context of short-wavelength perturbations where the metric perturbations can be neglected.

Importantly, this means that shear and bulk viscosity cannot stabilize a star that is unstable to radial perturbations. Imagine a configuration that is slightly too dense to be stable in the absence of viscosity, such as a supramassive neutron star whose mass is higher than the maximum ``TOV'' mass that defines the threshold of stability. In the absence of viscosity, this configuration is unstable to gravitational collapse, and if it is chosen as initial data, then the final state is a black hole. Now imagine adding an arbitrary amount of viscosity to this supramassive neutron star. Can the viscosity we just added to the fluid restore stability? Our analysis above suggests that this is never the case for the following reason. 

Let us think physically about the difference in the forces caused by the pressure of the star and the viscous forces.
In the perfect fluid case, the pressure --caused by conservative forces-- stops the collapse of the star, but viscosity is a dissipative force that \textit{slows down} the collapse of the star, yet it cannot stop it.
The viscous forces scale like gradients of the velocity, and thus, as the collapse of the star slows down, the viscous forces also decrease in magnitude, and they can never actually stop the star from collapsing.
Although this argument would require one to know the solution far away from the linear regime, where our results are valid, it is a useful physical interpretation of our results to explain why viscosity cannot radially stabilize an unstable star.


\subsubsection{Non-zero heat conductivity}\label{sssec:heat}

The stability analysis for heat conductivity based on Eq.~\eqref{eq:evcorr} is more complicated.
This is because the operator $\mathbb G$ is fourth order in radial derivatives and is not self-adjoint (and neither is the heat conductivity contribution to $\mathbb F$).
For this reason, we leave most of the details of this calculation to App.~\ref{app:fourth}.
The main result is that, given that the boundary terms are zero, $\mathbb G$ can be separated into a self-adjoint part $\mathbb S$ plus a skew-adjoint part $\mathbb A$.
From Eq.~\eqref{eq:evcorr}, only the self-adjoint part of the operator contributes to the analysis of stability, while the skew-adjoint part changes the frequencies of oscillations.
In the following, we focus on the self-adjoint part $\mathbb S$ only, since this is what determines stability, and 
discuss how to calculate 
the new frequencies in App. \ref{app:fourth}.

For vanishing boundary terms, the self-adjoint part of a fourth order differential operator can be written as
\begin{equation}
    W(r)\mathbb S\phi= \partial_r^2\left(L(r)\partial_r^2\phi\right)-\partial_r(M(r)\partial_r\phi)+N(r)\phi \ .
\end{equation}
Then, the expectation value of $\mathbb S$ is
\begin{align}
    \braket{\phi_j^{(0)}}{\mathbb S\phi_j^{(0)}}=\int_0^Rdr\, &L\left|\partial_r^2\phi_j^{(0)}\right|^2+M\left|\partial_r\phi_j^{(0)}\right|^2 \nonumber\\
    +&N\left|\phi_j^{(0)}\right|^2 \ ,
\end{align}
after integrating by parts a few times and neglecting boundary terms. 
Similarly, the contribution from the heat conductivity to the operator $\mathbb F$ can be separated into a self-adjoint and a skew-adjoint part.
This time, the operator only has two terms contributing to it, defined by some functions $K(r)$ and $H(r)$, which we present in detail in the appendix. 

With this in hand, we can now compute the corrections to the frequency. We focus on the purely imaginary part of the frequency correction, because if this is positive, then the modes are exponentially decaying, and thus, stable. Equivalently, we can also focus on the purely real part of the product $(i \omega_j^{(1)})$, which leads to stable modes if negative. With this in mind,  if we limit ourselves only to the contribution to $\omega_j^{(1)}$ coming from the heat conductivity, we find from Eq.~\eqref{eq:evcorr} that
\begin{equation}\label{eq:heatint}
\begin{split}
    \Re(i\omega_j^{(1)})= -&\frac{\int_0^Rdr\, L\left|\partial_r^2\phi_j^{(0)}\right|^2+M\left|\partial_r\phi_j^{(0)}\right|^2+N\left|\phi_j^{(0)}\right|^2}{\int_0^Rdr\, P\left|\partial_r\phi_j^{(0)}\right|^2+Q\left|\phi_j^{(0)}\right|^2}\\
    -&\frac{\int_0^Rdr\, K\left|\partial_r\phi_j^{(0)}\right|^2+H\left|\phi_j^{(0)}\right|^2}{\int_0^Rdr\, W\left|\phi_j^{(0)}\right|^2} \ ,
\end{split}
\end{equation}
where the denominator in the first line comes from recalling that
\begin{equation}
    \left(\omega_j^{(0)}\right)^2=\frac{\int_0^R dr\, P\left|\partial_r\phi_j^{(0)}\right|^2+Q\left|\phi_j^{(0)}\right|^2}{\int_0^R dr\, W\left|\phi_j^{(0)}\right|^2} \ .
\end{equation}
Then, stability to heat conductivity requires that Eq.~\eqref{eq:heatint} be negative, which, in turn, is determined by the integrals of the functions $L,M,N,K$ and $H$.
That is, a configuration is radial-mode stable to small heat conductivity (up to first order) if and only if the right-hand side of Eq.~\eqref{eq:heatint} remains negative for all eigenvectors $\phi_j^{(0)}$ of the perfect-fluid problem.
We note that if said integrals are only slightly positive, the bulk and shear viscosity can act against the heat conductivity to stabilize the star, in principle.

Even though Eq.~\eqref{eq:heatint} gives us a \textit{sufficient and necessary} condition for radial-mode stability, evaluating all the integrals in Eq.~\eqref{eq:heatint} for all eigenvectors is extremely complicated, and thus, we now search for simpler stability conditions.
For example, a sufficient condition for stability is for all of $L,M,N,K$ and $H$ to be positive everywhere, but this condition is very restrictive and is unlikely to be useful.
We can instead look for \textit{only necessary} conditions for stability based on high-frequency modes.
One should note that any fluid model breaks down at high enough frequencies, as one exits the regime of validity or cut-off scale of the effective field theory; however, the mode at which the model breaks down should be of the order of the ratio of the radius of the star to the mean free path.
Since this ratio is much larger than unity, we expect the condition we derive below to apply to modes before the cutoff frequency of the theory.

High-frequency modes are dominated by $|\partial_r^2\phi_j^{(0)}|\gg|\partial_r\phi_j^{(0)}|\gg|\phi_j^{(0)}|$.
Therefore, the terms that dominate in Eq.~\eqref{eq:heatint} are 
\begin{equation}\label{eq:highmodes}
    \Re\left(i\omega_j^{(1)}\right)\sim -\frac{\int_0^Rdr\, L\left|\partial_r^2\phi_j^{(0)}\right|^2}{\int_0^Rdr\, P\left|\partial_r\phi_j^{(0)}\right|^2}-\frac{\int_0^Rdr\, K\left|\partial_r\phi_j^{(0)}\right|^2}{\int_0^Rdr\, W\left|\phi_j^{(0)}\right|^2} \ .
\end{equation}
A necessary condition for stability based on high-frequency modes is then for the right-hand side of the above equation to be negative, which, in turn, depends on the relation between the first and second derivative of the eigenvectors. 
For example, if $|\partial_r^2\phi_j^{(0)}|/|\partial_r\phi_j^{(0)}|\gg|\partial_r\phi_j^{(0)}|/|\phi_j^{(0)}|$, then the first term dominates and $L(r)\geq 0$ is the sole condition needed.
Meanwhile, if $|\partial_r^2\phi_j^{(0)}|/|\partial_r\phi_j^{(0)}|\ll|\partial_r\phi_j^{(0)}|/|\phi_j^{(0)}|$, the second term dominates, and instead $K(r)\geq 0$ is the sole necessary condition.
In contrast, if the two ratios are comparable, we might require both $L(r)\geq0$ and $K(r)\geq 0$; however, this does not take into account that the two terms can counteract each other.
For example, if there is a region such that $K\leq 0$, then we should expect the high-frequency modes to be stable provided that $L\geq 0$ and $|L| \geq |K|$.

With this in mind, let us analyze the different cases described above. First, a sufficient condition for stability of high-frequency modes is  $L(r)\geq 0$ and $K(r)\geq 0$ everywhere inside the star. Conversely, a sufficient condition for instability of high-frequency modes is $L(r)\leq 0$ and $K(r)\leq 0$ everywhere inside the star.
Therefore, a necessary condition for stability is that $L(r)$ and $K(r)$ cannot both be negative everywhere inside the star.

We have shown that if $[L(r), K(r)]>0$ ($[L(r), K(r)]<0$) everywhere inside the star then the high-frequency modes will be stable (unstable), but what if neither of these conditions are met? That is, what happens if $L(r)\geq 0$ and $K(r)\leq 0$ in the star or viceversa?
In such cases, it is possible for the integral with $L(r)$ to stabilize the star against the contributions from the integral of $K(r)$ or viceversa. For these intermediate cases, we propose the following sufficient stability criterion for high-frequency modes. Let us assume that $|\partial_r^2\phi_j^{(0)}|/|\partial_r\phi_j^{(0)}|=|\partial_r\phi_j^{(0)}|/|\phi_j^{(0)}|=k$, one of whose solutions is $\phi_j^{(0)} = A_j \exp(i k r)$.
In this case, we can use Eq.~\eqref{eq:highmodes} to put a bound on the frequencies via
\begin{align}
    \Re\left(i\omega_j^{(1)}\right)\lesssim -k^2\Bigg[\ \ &\frac{\int_0^Rdr\, \inf(L/P)P\left|\partial_r\phi_j^{(0)}\right|^2}{\int_0^Rdr\, P\left|\partial_r\phi_j^{(0)}\right|^2}. \nonumber\\
    +&\frac{\int_0^Rdr\, \inf(K/W)W\left|\phi_j^{(0)}\right|^2}{\int_0^Rdr\, W\left|\phi_j^{(0)}\right|^2}\Bigg] \leq 0\ ,
\end{align}
where both infima are taken over the interval $(0,R)$.
The infima are simply numbers, and thus, they can be taken out of the integrals.
Now, suppose that we have an unstable configuration; then, $0\leq \Re(i\omega_j^{(1)})$, and seeing that $-k^2$ is always negative, we obtain the following \textit{sufficient} condition for the stability of high-frequency radial modes:
\begin{equation}\label{eq:condi}
    \inf\left(\frac LP\right)+\inf\left(\frac KW\right)\geq 0 \ .
\end{equation}
We remind the reader that this is \textit{not} a sufficient condition for the stability of the star as a whole, but rather just for the high-frequency modes.
If one finds that Eq.~\eqref{eq:condi} holds for some configuration, one must still check that the low frequency modes are stable.
However, it should suffice to check the first few low-frequency modes, and thus, such a stability check is much simpler than using all of Eq.~\eqref{eq:heatint} to check infinitely many modes.


Let us see what each of these conditions really implies at a physical level. 
Explicitly, the functions $L$ and $K$ are given by
\begin{subequations}
    \begin{align}
        L=&-\kappa \left(\frac{\varepsilon+p}{n}\right)^2\frac{e^{-\Lambda-4\Phi}}{r^2}c_n^2\left[c_s^2-c_n^2\right] \ ,\\
        K=&\kappa\left(\frac{\varepsilon+p}{n}\right)^2\frac{e^{\Lambda-2\Phi}}{r^2}\left[c_s^2-c_n^2\right] \ .
    \end{align}
\end{subequations}
Based on the discussion above and assuming that $\kappa\geq 0$, the condition $L\geq 0$ requires $c_n^2[c_n^2-c_s^2]\geq 0$, while the condition $K\geq 0$ requires $c_s^2\geq c_n^2$.
Let us now use that $c_s^2\in[0,1]$, which is a causality and stability condition for a perfect fluid.
Imposing this constraint, $L\geq 0$ can only occur if $c_n^2\leq 0$ or if $c_n^2\geq c_s^2\geq0$. Summarizing these conditions, we then have that
\begin{itemize}
    \item $K \geq 0$ when $c_n^2$ is smaller than or equal to $c_s^2$. 
    \item $L \geq 0$ when $c_n^2$ is negative or when $c_n^2$ is positive and greater than or equal to $c_s^2$.
\end{itemize}

Given the above conditions, we can now refine our sufficient and necessary conditions for stability. 
Let us begin by considering the sufficient condition discussed below Eq.~\eqref{eq:highmodes} (i.e. that $K(r)\geq 0$ and $L(r)\geq 0$). 
This condition, together with $c_s^2\in[0,1]$, implies that $c_n^2\leq 0$ inside the whole star is a sufficient condition for the high-frequency modes to be stable (because if $c_n^2$ is negative and $c_s^2$ is positive, then it is always true that $c_n^2 < c_s^2$). 
Let us now consider the necessary condition that $K(r)\geq 0$ or $L(r)\geq 0$ somewhere inside the star.
Using this condition, together with $c_s^2\in[0,1]$, implies that $c_s^2\geq c_n^2$ or $c_n^2\geq c_s^2$, which is always true.
Therefore, this necessary condition is automatically satisfied as long as $c_s^2\in[0,1]$.


To get an intuition for the conditions, we evaluate $c_n^2$ and $c_s^2-c_n^2$ for the Mathews equation of state of a relativistic ideal gas with particle mass $m$ \cite{Mathews1971}, given by
\begin{equation}
    p_{\rm Mathews}(\varepsilon,n)=\frac13\frac{\varepsilon^2-(mn)^2}{\varepsilon}.
\end{equation}
We find that
\begin{subequations}
\begin{align}
    c_n^2&:= \left(\frac{\partial p}{\partial \varepsilon}\right)_n = \frac13\left(\frac{mn}{\varepsilon}\right)^2\geq 0 \ ,\\
    \quad c_s^2-c_n^2&=\frac{n}{\varepsilon+p}\frac{\partial p}{\partial n}\eval_\varepsilon=-\frac23\frac{n}{\varepsilon+p}\left(\frac{mn}{\varepsilon}\right)m\leq 0 \ ,
\end{align}
\end{subequations}
where the first equality in the second line is a known thermodynamic identity (see, for example, Eq.~$(19)$ in \cite{Bemfica2022}).
Therefore, $L\geq 0$ but $K \leq 0$ for the Mathews ideal gas, which means the sufficient condition for stability discussed above is not satisfied, although the necessary one is (because it is always satisfied, as shown above). To show whether this equation of state leads to stable high-frequency modes we must either compute all (the infinite number) of them explicitly, or check Eq.~\eqref{eq:condi} for a finite set of them to determine if the infima-form of the sufficient condition is satisfied.
Recall that an instability in the term proportional to $K$ may be counteracted by bulk and shear viscosity terms, by the results of Sec.~\ref{sssec:bulkshear}.
In fact, with enough bulk and/or shear viscosity, it is always possible to counteract the type of instability to heat conductivity we describe in this work.

In \cite{Lindblom1983}, Hiscock and Lindblom obtained a stability criterion for an Eckart fluid based on energy arguments.
They argue that, if the energy associated with the viscous perturbations is greater than that of the energy associated with perfect fluid contributions, then the star will be unstable.
In particular, they find that for high-frequency modes, contributions from the heat conductivity always exceed those of a perfect fluid.
However, since the hydrodynamic theory breaks down at small scales, there is a highest possible frequency admitted by hydrodynamics, and as long as the highest frequency is stable, no such instability can occur.
This effectively sets an upper bound on the value of the heat conductivity.

The condition of \cite{Lindblom1983} is similar to ours in that it is based on high-frequency modes associated with the heat conductivity.
However, our condition in Eq.~\eqref{eq:condi} (and the discussion that followed) is not equivalent to that of \cite{Lindblom1983}.
For instance, we assume small viscosity and thus, in principle, the contribution from the heat conductivity is always smaller than perfect fluid contributions.
In fact, the value of $\kappa$ does not affect our stability criteria (beyond that, if it is small enough, an instability can be counteracted by bulk and shear viscosity), and thus, no bound on the heat conductivity (such as that given in \cite{Lindblom1983}) can be derived from our analysis.
Moreover, in our analysis, the heat conductivity does not dominate for high-frequency modes when compared to the contribution from bulk and shear viscosity (unless $|\partial_r^2\phi_j^{(0)}|/|\partial_r\phi_j^{(0)}|\gg|\partial_r\phi_j^{(0)}|/|\phi_j^{(0)}|$ is satisfied).
Therefore, our condition is independent of that in \cite{Lindblom1983}, and we, in fact, do not recover the same condition as in that paper.
We note that the condition derived in \cite{Lindblom1983} is based on the heat current of Eq.~\eqref{eq:oeck} rather than Eq.~\eqref{eq:eckq}, which is what we used in our derivation.
This might be an explanation for why the conditions are different, and it would require a rederivation of our analysis with Eq.~\eqref{eq:oeck}, or of their analysis with Eq.~\eqref{eq:eckq} for a proper comparison.


\subsection{BDNK Fluid}

We now proceed to carry out the small viscous coefficient expansion for a BDNK fluid by rescaling each viscous coefficient $(\zeta,\eta,\kappa,\tau_{\mathscr E},\tau_{\mathscr P},\tau_{\mathscr Q})$ by a small parameter $\alpha$,
and then expanding Eqs.~\eqref{eq:psi}, \eqref{eq:ds}, \eqref{eq:de}, \eqref{eq:dq} and \eqref{eq:biggg} in the small parameter.
Scaling all coefficients by the same parameter ensures that, if the causality conditions of the model [see Eq.~(21) of \cite{Bemfica2022}] are satisfied before scaling, they will also be satisfied after scaling.
However, this scaling causes $\Delta\mathscr Q_r$ and $\Delta\mathscr E$ to also become scaled by a factor of $\alpha$, which results in a singular perturbation problem. Physically, this occurs because several scales arise and separate in the small viscosity limit.
This separation becomes apparent by looking at the equations of motion, and seeing that Eqs.~\eqref{eq:psi}, \eqref{eq:ds}, and \eqref{eq:biggg} behave as
\begin{equation}
    \frac{\dot\Xi}{\partial_t\dot\Xi}\sim \frac{1}{\eta}\sim \frac{1}{\alpha},
\end{equation}
while Eqs.~\eqref{eq:de} and \eqref{eq:dq} behave as
\begin{equation}
    \frac{\Delta\mathscr E}{\partial_t\Delta\mathscr E}\sim \tau_{\mathscr E}\sim \alpha \ .
\end{equation}
In other words, when we introduce small viscosity, we have two types of variables of motion: 
\begin{itemize}
\item Those that evolve slowly, as $\dot\Xi/\partial_t\dot\Xi\sim \alpha^{-1}$, and thus, depend on a ``slow'' or ``long'' timescale (which is the same as that of the non-viscous perturbations), 
\item Those that evolve fast, as $\Delta\mathscr E/\partial_t\Delta\mathscr E\sim \alpha$, and thus, depend on a ``fast'' or ``short'' timescale.
\end{itemize}
Such a system begs for the use of multiple-scale analysis, in which one introduces both timescales into the ansatz, to search for a solution. Notice that the appearance of multiple scales occurs because of the introduction of relaxation times in the constitutive relations; therefore, an Eckart fluid does not present such behavior, and it does not require multiple-scale analysis.

Contrary to the usual multiple-scale analysis used in gravitational wave analysis during the inspiral phase (see, for example, \cite{PoissonWill}), however, the resulting equations also require a radial variable rescaling.
This happens because Eqs.~\eqref{eq:de} and \eqref{eq:dq} have the following scaling of the radial derivatives:
\begin{equation}\label{eq:rscale}
    \frac{\Delta\mathscr E}{\partial_r\Delta\mathscr Q_r}\sim \tau_{\mathscr E}\sim \alpha,\quad \frac{\Delta\mathscr Q_r}{\partial_r\Delta\mathscr E}\sim \frac{\tau_\mathscr P}{\tau_{\mathscr E}}\tau_{\mathscr Q}\sim \alpha \ .
\end{equation}
Since $\Delta\mathscr E$ and $\Delta\mathscr Q_r$ have the same scaling\footnote{Taking a radial derivative of the second equation in Eq.~\eqref{eq:rscale} and substituting into the first equation results in $\Delta\mathscr E/(\partial_r^2\Delta\mathscr E)\sim \tau_{\mathscr E}\tau_{\mathscr Q}\sim \alpha^2$, leading to the same result.}, it follows that $\Delta\mathscr E/(\partial_r\Delta\mathscr E)\sim \alpha$ and similarly for $\Delta\mathscr Q_r$.

All of this, then, naturally suggests the use of multiple scale analysis in both temporal and radial scales.
By the analysis above, we must introduce two timescales and two radial scales, and then our solution will be of the form
\begin{equation}
    \Xi=\Xi(t,\tau,r,\rho) \ ,
\end{equation}
where $\tau\equiv t/\alpha$ and $\rho\equiv r/\alpha$ are the short/fast timescale and the short radial scale, respectively\footnote{More accurately, $\tau$ is the time coordinate rescaled such that fast processes take finite time even in the $\alpha\to 0$ limit, and $\rho$ having an equivalent definition but for the radial coordinate. We will refer to them as the fast timescale and short radial scale as is usually done in the literature of multiple scale analysis e.g.~\cite{PoissonWill}.}.
The field $\Xi(t,\tau,r,\rho)$ is then expanded perturbatively and secular terms are eliminated.
The dependence of the solution on the fast timescale will exhibit exponential decay, which leads to the overall solution being the sum of a part that depends only on the slow timescale, plus another part that depends only on the fast timescale.
In theory, the solution of the fast timescale could be modulated by a slowly-varying amplitude, but this would result in secular terms, unless the amplitude has no dependence on the slow timescale.

With all of these considerations in mind, our solution should be of the form
\begin{equation}
    \Xi(t,\tau,r,\rho)=\Xi(t,r,\rho)+\Tilde\Xi(\tau,r,\rho) \ .
\end{equation}
Such a splitting is well established in the case of ordinary differential equations and results in the O'Malley-Vasil'eva expansion \cite{OMalley1971,Vasileva1963,Verhulst}.
We highlight that the reason for this splitting is the exponential decay of the terms that are involved in the fast timescale, contrary to the usual approach of multiple timescales for oscillatory behavior in the faster scale.
Even though we are working with partial differential equations, said result also applies to our problem.
We will not carry out the computation here for brevity, but it follows through a similar calculation to those mentioned above.
Moreover, this multiple-scale analysis is equivalent to carrying out a boundary layer analysis, where the domain inside the boundary layer corresponds to the fast timescale (occurring for small times), while the slow timescale corresponds to the domain outside the boundary layer.

The solution is then the sum of a fast timescale part plus a slow timescale part, which we explicitly write out as
\begin{equation}\label{eq:multiscale}
    \Xi(t,r)=\sum_{j\geq 0}\alpha^j\Xi^{(j)}(t,r)+\sum_{j\geq 0}\alpha^j\Tilde\Xi^{(j)}(t/\alpha,r,r/\alpha) \ .
\end{equation}
In this notation, $\Xi^{(j)}$ denotes the solution at order $j$ of the perturbation in the \textit{slow} timescale, 
while tilde variables (i.e. $\Tilde\Xi^{(j)}$) denote solutions in the \textit{fast} timescale.
We will call the first sum in Eq.~\eqref{eq:multiscale} the slow timescale solution, while the second sum we call the fast timescale solution.
Moreover, the analysis for ordinary differential equations suggests that those variables that evolve in the slow timescale are zero at order zero in the fast timescale, i.e. $\Tilde\Xi^{(0)}=\Tilde{\dot\Xi}^{(0)}=\Delta\Tilde s^{(0)}=\Tilde\Psi^{(0)}=0$, but $\Delta\Tilde {\mathscr E}^{(0)}$ and $\Delta\Tilde{\mathscr Q}_r^{(0)}$ are both nonzero, and we will make such assumptions in our analysis.
We will analyze the slow timescale in Sec.~\ref{sssec:slow} and the fast timescale in Sec.~\ref{sssec:fast}.

\subsubsection{Slow timescale evolution}\label{sssec:slow}

In the slow timescale, since $\Delta\mathscr Q_r$ and $\Delta\mathscr E$ are at least order $\alpha$, we can set $\Delta\mathscr Q_r^{(0)}=0,\Delta\mathscr E^{(0)}=0$.
By Eq.~\eqref{eq:psi}, this leads to $\partial_t\Psi^{(0)}=0$, while at order zero, Eq.~\eqref{eq:de} results in $\Delta\mathscr E^{(1)}=0$, and thus in Eq.~\eqref{eq:ds}, we find $\partial_t\Delta s^{(0)}=0$.
Just as we did at order zero for an Eckart fluid, we can then set $\Delta s^{(0)}=0$ and $\Psi^{(0)}=0$.
Meanwhile, from Eq.~\eqref{eq:dq}, it follows that: 
\begin{equation}\label{eq:q1}
    \Delta\mathscr Q_r^{(1)}=\kappa T\frac{\varepsilon+p}{n}\partial_r\delta\varphi^{(0)} \ ,
\end{equation}
so that once we substitute everything into Eq.~\eqref{eq:biggg}, the only nonzero terms come from the first line.
Therefore, at order zero, we once again recover the perfect-fluid problem of  Eq.~\eqref{eq:chandra}.

Moving on to first order, using $\partial_t\Delta\mathscr E^{(1)}=0$, Eqs.~\eqref{eq:ds} and \eqref{eq:de} combine into
\begin{equation}
    nT\partial_t\Delta s^{(1)}=-\frac{e^{-\Lambda}}{r^2}\partial_r\left(e^{-\Lambda-2\Phi}r^2\Delta\mathscr Q_r^{(1)}\right) \ , 
\end{equation}
while Eq.~\eqref{eq:psi} results in
\begin{equation}
    \partial_t\Psi^{(1)}=-4\pi re^{-\Phi}\Delta\mathscr Q_r^{(1)} \ .
\end{equation}
Once we substitute Eq.~\eqref{eq:q1}, we obtain the same expressions for $\partial_t\Delta s^{(1)}$ and $\partial_t\Psi^{(1)}$ as we found in Sec. \ref{ssec:smalleck} for the Eckart fluid.
Therefore, $\Psi^{(1)}$ and $\Delta s^{(1)}$ are given by Eqs.~\eqref{eq:psiexp} and \eqref{eq:sexp} for the BDNK fluid in the slow timescale just like for the Eckart fluid.
Meanwhile, Eq.~\eqref{eq:dq} when expanded at first order reduces to
\begin{align}
    &\frac{e^{-\Phi}}{\tau_{\mathscr Q}}\left[\Delta \mathscr Q_r^{(2)}-\kappa T\frac{\varepsilon+p}{n}\partial_r\delta\varphi^{(1)}\right] \nonumber\\
    &\quad \quad\quad\quad=\partial_r\left[(\zeta+\tfrac43\eta)\left(\frac{e^{-\Phi}}{r^2}\partial_r\dot\Xi^{(0)}\right)\right] \nonumber \\
    &\quad\quad\quad\quad \quad\ +\frac{4\eta}{r}\left[\frac{e^{-\Phi}}{r^2}\Phi'\dot\Xi^{(0)}\right] -\partial_t\Delta\mathscr Q_r^{(1)}\ .
\end{align}
Once we substitute the above into Eq.~\eqref{eq:biggg}, we once again obtain Eq.~\eqref{eq:exp1}.
Then, since $\Psi^{(1)}$ and $\Delta s^{(1)}$ are the same as for the Eckart problem, we find that $\Xi^{(1)}$ also has the same solution as it did for the Eckart problem.
Therefore, the stability analysis of Sec. \ref{ssec:smalleck} also applies to a BDNK fluid in the slow timescale.
This means that for long timescales, the BDNK and Eckhart fluids have the same modes when viscosity is sufficiently small.
However, we note that at next order in small viscous parameter, the equations for a BDNK and an Eckart fluid do differ in the slow timescale, and thus, they should lead to different corrections to the eigenfrequencies. This, however, is beyond the scope of this paper.

\subsubsection{Fast timescale evolution}\label{sssec:fast}

We now move on to analyze the fast timescale of the problem, i.e. the second sum in the expansion Eq.~\eqref{eq:multiscale}.
Substituting in the ansatz and solving order by order, 
%
%
Eqs.~\eqref{eq:de} and \eqref{eq:dq} 
lead to
\begin{subequations}\label{eq:tele}
\begin{align}
    \partial_\tau\Delta\Tilde{\mathscr E}^{(0)}=-\frac{e^{-\Phi}}{\tau_{\mathscr E}}\Delta \Tilde{\mathscr E}^{(0)} -e^{-2\Lambda-\Phi}\partial_\rho\Delta\Tilde{\mathscr Q}_r^{(0)}\ , \label{eq:telee}\\
    \partial_\tau\Delta\Tilde{\mathscr Q}_r^{(0)}=-\frac{e^{-\Phi}}{\tau_{\mathscr Q}}\Delta \Tilde{\mathscr Q}_r^{(0)}-e^{-\Phi}\frac{\tau_\mathscr P}{\tau_\mathscr E}\partial_\rho\Delta\Tilde{\mathscr E}^{(0)} \label{eq:teleq}\ .
\end{align}
\end{subequations}
Notice that the coefficient functions (i.e.~$\Phi$, $\Lambda$) depend only on the slow scale $r$, and thus there is no explicit $\rho$ dependence in the equation. 
Therefore, we can consider the above equation as having constant coefficients, so we will suppress the explicit dependence on $r$ for the sake of brevity.
Physically, we can think of this result as choosing a spherical shell of radius $r$, zooming into it, and looking at changes in the small radial scale $\rho$.

The system decouples by taking a time derivative of Eq.~\eqref{eq:telee} and a radial derivative of Eq.~\eqref{eq:teleq}. Doing so, the equations can be combined into a single, second-order wave equation with dissipation, specifically
\begin{align}\label{eq:telefull}
    &\left[\partial_\tau^2-e^{-2\Lambda-2\Phi}\frac{\tau_\mathscr P}{\tau_\mathscr E}\partial_\rho^2\right]\Delta\Tilde{\mathscr E}^{(0)} \nonumber\\
    &+e^{-\Phi}\left[\frac{1}{\tau_\mathscr E}+\frac{1}{\tau_\mathscr Q}\right]\partial_\tau\Delta\Tilde{\mathscr E}^{(0)}+\frac{e^{-2\Phi}}{\tau_\mathscr E\tau_\mathscr Q}\Delta\Tilde{\mathscr E}^{(0)}=0 \ .
\end{align}
The other variable $\Tilde{\mathscr Q}_r^{(0)}$ satisfies the same equation as Eq.~\eqref{eq:telefull}, which can be obtained by combining a radial derivative of Eq.~\eqref{eq:telee} and a time derivative of Eq.~\eqref{eq:teleq}.

These equations can be solved via separation of variables, which will lead to a harmonic decomposition $\Delta\Tilde{\mathscr E}^{(0)}(\tau,\rho)=e^{i\Omega\tau}\psi_{\mathscr E}(\rho)$,
and likewise for $\Delta\Tilde{\mathscr Q}_r^{(0)}(\tau,\rho)=e^{i\Omega\tau}\psi_{\mathscr Q}(\rho)$.
At this point, we obtain an eigenvalue problem for the second derivative and we must impose boundary conditions.
We choose to impose $\partial_\rho\psi_{\mathscr E}(0)=0$ and $\psi_{\mathscr E}(R/\alpha)=0$, 
using the fact that $\mathscr E$ is a scalar variable, and hence, is an even function about the center of the star, 
and that the correction to the energy density should be zero at the surface of the star.
By Eq.~\eqref{eq:tele}, the boundary conditions on $\psi_{\mathscr E}$ imply the following boundary conditions for $\psi_{\mathscr Q}$: $\psi_{\mathscr Q}(0)=0$ and $\partial_\rho\psi_{\mathscr Q}(R/\alpha)=0$.
The resulting eigenfunctions and eigenvalues are then
\begin{subequations}
\begin{align}
    \psi_{j,\mathscr E}(\rho)&=\cos(\sqrt{\lambda_j}\rho) \ ,\\
    \psi_{j,\mathscr Q}(\rho)&=ie^{-2\Lambda+\Phi}\frac{\tau_\mathscr E}{\tau_\mathscr P}\frac{e^{-\Phi}/\tau_\mathscr Q+i\Omega_j}{e^{-\Phi}/\tau_\mathscr E+i\Omega_j}\sin\left(\sqrt{\lambda_j}\rho\right) \ ,\\
    \lambda_j&=\left(\frac{(j+\tfrac12)\pi\alpha}{R}\right)^2,\quad j=0,1,\dots \ .
\end{align}
\end{subequations}

The system written in the form of Eq.~\eqref{eq:telefull} is hyperbolic with a finite speed of propagation, given by $e^{-\Lambda-\Phi}\sqrt{\tau_\mathscr P/\tau_\mathscr E}$.
We can compare this with the speed of propagation of the Chandrasekhar problem [i.e.~Eqs.~\eqref{eq:chandra}, \eqref{eq:wpq}], which is given by $\sqrt{P/W}=e^{-\Lambda-\Phi}c_s$.
Following the classification of characteristics of BDNK in \cite{Disconzi2024}, the speed of propagation for the oscillations in the fast timescale is the \textit{speed of second sound}.
Seeing that the speed of propagation of perturbations for a perfect fluid also have a factor of $e^{-\Lambda-\Phi}$, we say that the speed of second sound is given by
\begin{equation}
    c_\tau^2=\frac{\tau_\mathscr P}{\tau_\mathscr E} \ .
\end{equation}
For the perturbations in the fast timescale to be causal, we must then require that $\tau_\mathscr E\geq \tau_\mathscr P$.
This is one of the causality conditions that BDNK already found, when taking $\zeta=0,\eta=0$, and $\kappa=0$ [see Eq.~(21d) of \cite{Bemfica2022}].
The ability to recover the causality conditions for a BDNK fluid by looking at neutron star perturbations was first noted in \cite{RedondoYuste2024}.
The reason why we only find the causality condition for the simplified system with $\zeta=0,\eta=0,$ and $\kappa=0$, 
is that, by considering only the fast timescale, we removed the contribution from the other coefficients.
This is similar to how, in contrast, the slow timescale was equivalent to just the Eckart problem without the relaxation times.
Thus, by separating the solution into two different timescale, we are effectively separating the problem into a solely Eckart (slow timescale) and solely ``corrections to Eckart'' (fast timescale) problem.
This separation only fully occurs at lowest order in viscosity, and at higher orders, $\zeta,\eta,$ and $\kappa$ will appear in the fast timescale, while $\tau_\mathscr E,\tau_\mathscr P,$ and $\tau_\mathscr Q$ will appear in the slow timescale.
Therefore, we expect the more complex behavior of a BDNK fluid, such as the additional causality conditions, to arise at higher order in perturbation theory.

We now move on to analyzing the eigenfrequencies.
From Eq.~\eqref{eq:telefull}, the mode expansion is equivalent to the transformation $\partial_\tau^2\to -\Omega_j^2$ and $\partial_\rho^2\to -\lambda_j$, so that we obtain a pair of eigenfrequencies for each eigenvalue.
Explicitly we obtain
\begin{align}\label{eq:fasteig}
    \Omega_j^\pm=&\frac i2e^{-\Phi}\left[\frac{1}{\tau_\mathscr E}+\frac{1}{\tau_\mathscr Q}\right] \nonumber\\
    &\pm e^{-\Phi}\sqrt{e^{-2\Lambda}c_\tau^2\lambda_j-\frac14\left[\frac{1}{\tau_\mathscr E}-\frac{1}{\tau_\mathscr Q}\right]^2} \ .
\end{align}
Then, the general solution to the problem in the fast timescale is given by 
\begin{align}
    \Delta\Tilde{\mathscr E}^{(0)}&=\sum_{j\geq 0}a_j^+e^{i\Omega_j^+\tau}\psi_{j,\mathscr E}^+(\rho)+a_j^-e^{i\Omega_j^-\tau}\psi_{j,\mathscr E}^-(\rho) \ ,\\
    \Delta\Tilde{\mathscr Q}_r^{(0)}&=\sum_{j\geq 0}a_j^+e^{i\Omega_j^+\tau}\psi_{j,\mathscr Q}^+(\rho)+a_j^-e^{i\Omega_j^-\tau}\psi_{j,\mathscr Q}^-(\rho) \ ,
\end{align}
where the $a_j^\pm$ (and likewise $\Omega_j^\pm$) are understood to be functions of $r$.
One might be concerned with the possibility that the term inside the square root of Eq.~\eqref{eq:fasteig} might lead to the imaginary part of $\Omega_j^\pm$ becoming negative for some eigenvalues.
In particular, since the rescaling parameter $\alpha$ is arbitrary in our expansion, the eigenvalues can be made arbitrarily small.
However, taking the limit ${\alpha}/{R}\to 0$ corresponds to making the domain (semi-)infinite, which would lead to a continuous spectrum of modes $\kappa^2\in(0,\infty)$ with their corresponding eigenfrequency.
However, because $\lambda_j>0$, the most unstable possible modes correspond to $\lambda_j\to 0$ (which is never a ``true'' mode unless we take the limit $\alpha/R\to 0$).
Therefore, as long as the modes resulting from taking the limit $\lambda_j\to 0$ are stable, any configuration is stable. 
Explicitly, we have
\begin{equation}
    \lim_{\alpha/R\to 0}\Omega_0^\pm=\frac{i}{2}e^{-\Phi}\left[\frac{1}{\tau_\mathscr E}+\frac{1}{\tau_\mathscr Q}\right]\pm \frac i2e^{-\Phi}\left|\frac{1}{\tau_\mathscr E}-\frac{1}{\tau_\mathscr Q}\right| \ ,
\end{equation}
and thus, the most possibly unstable modes are given by
\begin{equation}\label{eq:fulldecay}
    \lim_{\alpha/R\to 0}\Omega_0^+=i\frac{e^{-\Phi}}{\tau_\mathscr E},\quad \lim_{\alpha/R\to 0}\Omega_0^-=i\frac{e^{-\Phi}}{\tau_\mathscr Q} \ ,
\end{equation}
where the choice of which mode is the $+$ or $-$ mode depends on whether $\tau_\mathscr E>\tau_\mathscr Q$ or viceversa,
with the above being the choice when $\tau_\mathscr Q>\tau_\mathscr E$.
Therefore, one recovers the conditions $\tau_\mathscr E>0$ and $\tau_\mathscr Q>0$ for stability.
Notice that $\tau_\mathscr E>0$ and $c_\tau^2>0$ then imply that $\tau_\mathscr P>0$ and thus we also recover the positivity of the relaxation times by requiring stability and causality of the solution in the fast timescale.
We highlight that the two decaying modes given by Eq.~\eqref{eq:fulldecay} would be the only two modes found if we had \textit{not} rescaled the radial variable.
Importantly, if the radial variable is not rescaled together with time, we do not obtain the Telegrapher equations [Eq.~\eqref{eq:tele}], and thus, there would be no finite propagation associated with the speed of second sound.
In the next section, we will actually see that this behavior is \textit{not} found in the case of the Maxwell-Cattaneo equations.

Before proceeding, we would like to comment on the corrections at next order for the expansion in the fast timescale.
The resulting equations at first order will be the same pair of equations [Eq.~\eqref{eq:tele}] as those at order zero but with a source term composed of an off-diagonal multiplication operator of the solution at order zero.
Since $\sin(\lambda\rho)$ and $\cos(\lambda\rho)$ are orthogonal to each other, this does not lead to correction to the eigenvalues, but rather to a mixing of modes.
Moreover, due to the source term and boundary conditions, the solution will have terms polynomial in $\tau$ and $\rho$.
These terms do not lead to an instability because they are multiplied by $e^{i\Omega\tau}$, whose decaying behavior supersedes polynomial growth.
However, when reconstructing the solution by substituting $\tau=t/\alpha$ and $\rho=r/\alpha$, this will lead to terms in the series of $\alpha$ that appear at lower order than expected.
Nevertheless, the series, while using the variables $\tau$ and $\rho$ is well-defined, and thus, we expect that when carried out to all orders, the series can be resummed in the $\tau$ and $\rho$ variables before scaling back to $\tau=t/\alpha,\rho=r/\alpha$, in which case the solution will be well-behaved.
The fact that the series is not well-behaved is not surprising since terms of the form $e^{-t/\alpha}$ show up explicitly in our solution.
This type of behavior is associated with non-perturbative expansion, for which one should use superasymptotic/hyperasymptotic approximations or transseries to obtain a well-behaved solution \cite{Boyd2000}.
We leave the exploration of such methods to solve the perturbative problem for future work.

\subsection{Maxwell-Cattaneo Fluid}

We now carry out the same process as in the previous subsection but for the equations that result from using the Maxwell-Cattaneo equations [Eqs.~\eqref{eq:psi}, \eqref{eq:dseck}, \eqref{eq:misrel} and \eqref{eq:bigg3}].
Just like in the case of a BDNK fluid, the equations of motion require us to introduce an additional timescale to solve perturbatively.
For example, Eq.~\eqref{eq:misrel} has $\Delta\mathscr P/\partial_t\Delta\mathscr P\sim \tau_1\sim \alpha$, and $\Delta\mathscr Q_r$ and $\Delta\pi_r^r$ have a similar scaling.
Therefore, we need to introduce an additional timescale to carry out the perturbative analysis.
The two timescales are $t$ and $\tau\equiv t/\alpha$, and the analysis proceeds in a similar manner to what we did in the previous subsection.
However, unlike for the BDNK fluid, this time we do \textit{not} introduce a rescaled radial variable.
The reason for this is that Eqs.~\eqref{eq:misrel} do not include radial derivatives of the viscous fluxes on the right-hand side, and thus, there are no relations of the form of Eq.~\eqref{eq:rscale}, as in the case of a BDNK fluid.
This means that if we were to assume a $\rho\equiv r/\alpha$ dependence in the fields, these cannot possibly change with $\rho$, and thus, any dependence on $\rho$ would remain undetermined.
Therefore, the equations cannot lead to oscillatory behavior associated with the fast timescale, as we found for a BDNK fluid.
Following the Ansatz Eq.~\eqref{eq:multiscale}, we introduce variables $\Xi(t,r)$ and $\Tilde\Xi(\tau,r)$, which are expanded in a small parameter, and correspond to the slow and fast timescale solutions, respectively. 



Like before, we start by analyzing the slow timescale.
At zeroth order Eq.~\eqref{eq:misrel} tells us that $\Delta\mathscr P^{(0)}=0,\Delta\mathscr Q_r^{(0)}=0$, and $\Delta(\pi_r^r)^{(0)}=0$.
Once again, Eqs.~\eqref{eq:psi} and \eqref{eq:dseck} with appropriate initial condition result in $\Psi^{(0)}=0$ and $\Delta s^{(0)}=0$, and thus, Eq.~\eqref{eq:bigg3} at order zero
reduces to the Chandrasekhar equation [Eq.~\eqref{eq:chandra}], as expected.
We now move to the expansion at first order for the slow timescale.
First, Eqs.~\eqref{eq:misrel} become
\begin{subequations}
    \begin{align}
        \Delta\mathscr P^{(1)}&=-\frac{\zeta}{r^2}\partial_r\dot\Xi^{(0)} \ ,\\
        \Delta\mathscr{Q}_r^{(1)}&=\kappa T\frac{\varepsilon+p}{n}\partial_r\delta\varphi^{(0)} \ ,\\
        \Delta(\pi_r^r)^{(1)}&=-\frac43\frac{\eta}{r^2}\partial_r\dot\Xi^{(0)}+\frac{4\eta}{r}\frac{\dot\Xi^{(0)}}{r^2} \ .
    \end{align}
\end{subequations}
In particular, $\Delta\mathscr Q_r^{(1)}$ does not change from that of the Eckart fluid, while $\Psi^{(1)}$ and $\Delta s^{(1)}$ are governed by the same equations as before,
and are thus also given by Eqs.~\eqref{eq:psiexp} and \eqref{eq:sexp} respectively.
Finally, we must substitute into Eq.~\eqref{eq:bigg2} for which the only terms that can differ from the Eckart results are those in the last line.
We evaluate these to find
\begin{align}
    &-\partial_r\left[e^{\Lambda-2\Phi}\left(\Delta \mathscr P^{(1)}+\Delta(\pi_r^r)^{(1)}\right)\right]-\frac3re^{\Lambda-2\Phi}\Delta(\pi_r^r)^{(1)} \nonumber\\
    =&\partial_r\left[\frac{e^{\Lambda-2\Phi}}{r^2}\left(\zeta+\tfrac43\eta\right)\dot\Xi^{(0)}\right]-\frac{4\eta}{r}\frac{e^{\Lambda-2\Phi}}{r^2}(\Lambda'-2\Phi') \nonumber\\
    &\ -\frac{4\partial_r\eta}{r}\frac{e^{\Lambda-2\Phi}}{r^2}
\end{align}
Therefore, we obtain once again Eq.~\eqref{eq:exp1}, and thus, the slow timescale perturbations are exactly the same as the Eckart and BDNK perturbations to first order. We conclude that the stability analysis of Secs.~\ref{sssec:bulkshear} and \ref{sssec:heat} also hold for a MIS fluid.

We now move on to analyzing the fast timescale for the fluid.
Similar to before, $\Tilde\Xi^{(0)}=0,\Tilde{\dot\Xi}^{(0)}=0,\Delta\Tilde s^{(0)}=0,$ and $\Tilde\Psi^{(0)}=0$,
and thus, at zeroth order, only the left-hand side of Eq.~\eqref{eq:misrel} is nonzero.
This leads to relaxation-type equations with solutions
\begin{subequations}\label{eq:decayexp2}
    \begin{align}
        \Delta\Tilde{\mathscr P}^{(0)}(\tau,r)&=\Delta\Tilde{\mathscr P}^{(0)}(0,r)\exp\left(-\frac{e^{-\Phi}}{\tau_{0}}\tau\right) \ ,\\
        \Delta\Tilde{\mathscr Q}_r^{(0)}(\tau,r)&=\Delta\Tilde{\mathscr Q}_r^{(0)}(0,r)\exp\left(-\frac{e^{-\Phi}}{\tau_{1}}\tau\right) \ ,\\
        \Delta(\Tilde{\pi}_r^r)^{(0)}(\tau,r)&=\Delta(\Tilde{\pi}_1^1)^{(0)}(0,r)\exp\left(-\frac{e^{-\Phi}}{\tau_{ 2}}\tau\right) \ .
    \end{align}
\end{subequations}
Going to first order, one can check that $\Tilde\Xi^{(1)},\Tilde{\dot\Xi}^{(1)},\Tilde\Psi^{(1)},$ and $\Delta\Tilde s^{(1)}$ are simply sourced by the decaying modes of Eq.~\eqref{eq:decayexp2}.
Meanwhile, the equations at first order from Eq.~\eqref{eq:misrel} do not change from those at zeroth order,
except that Eqs.~\eqref{eq:misp} and \eqref{eq:mispi} will now be sourced by $\Delta\Tilde{\mathscr Q}_r^{(0)}$, and thus, we will start to see mode-mixing.
This can lead to secular-like terms, as we saw for the BDNK case, but this time, this only occurs when either $\tau_0=\tau_1$ or $\tau_2=\tau_1$; therefore, in the general case, the series expansion is regular (at least at first order in perturbation theory).
In terms of stability, this implies that, for the slow timescale, the analysis is the same as that of Eckart and BDNK fluids, carried out in Sec.~\ref{ssec:smalleck}.
Meanwhile, for the fast timescale, all modes are of the decaying type, according to Eq.~\eqref{eq:decayexp2}, and thus, solutions will always be stable (as long as one chooses $(\tau_1,\tau_2,\tau_3)>0$).
We conclude that the three fluids have the same stability properties for small viscosities.

\section{Conclusions}\label{sec:conclusions}

We have studied the stability of out-of-equilibrium neutron stars by analyzing radial perturbations for Eckart, BDNK, and MIS fluids in the small viscosity limit.
This is the first study of stability of neutron stars for BDNK fluids, and the first stability analysis based on the explicit calculation of modes for spherically-symmetric (i.e.~non-rotating) neutron stars for Eckart and MIS fluids.
Our analysis includes contributions from \textit{all} transport coefficients; that is, we take into account both bulk and shear viscosity, heat conductivity, and the relaxation behavior seen in both BDNK and MIS fluids, when appropriate.
Although our analysis is done for the case of small viscosity, we believe this is appropriate, since most neutron stars are expected to have small viscous effects compared to equilibrium effects \cite{Alford2018,Shternin2008,Chabanov2025,Most2022}.

In the small viscosity approximation, the perturbation equations can be solved using multiple-scale analysis.
For BDNK and MIS fluids, such an analysis separates the problem into two timescales, a fast (or short) scale and a slow (or long) scale.
Meanwhile, for Eckart fluids, only perturbations with a slow timescale exist.
Moreover, BDNK fluids also have a second, short radial scale that appears only in the fast timescale part of the solution, while no such radial rescaling is present for either MIS or Eckart fluids.
We show here that, at least to first order in small viscosity, all three fluids are governed by the same equations in the slow timescale,
and thus, they have the same (slow timescale) stability properties.
In this timescale, only the bulk and shear viscosities, and the heat conductivity contribute at first order.

We obtain stability conditions in the slow timescale against each of the three transport phenomena.
We find that a star will always be stable to bulk and shear viscosity, which is a consequence of the fact that a certain Sturm-Liouville operator is positive [Eq.~\eqref{eq:stablvis}].
Meanwhile, stability to heat conductivity is more complicated because it has contributions from two non-self-adjoint operators of order $2$ and $4$ in radial derivatives.
Only the symmetric parts of these operators contribute to stability, and, from the behavior of high-frequency modes, we can determine stability conditions to heat conduction.
The stability conditions involve the adiabatic speed of sound $c_s^2=\partial p/\partial\varepsilon\eval_s$, and a similar related quantity $c_n^2\equiv \partial p/\partial\varepsilon\eval_n$.
A necessary stability condition is for either $c_n^2[c_s^2-c_n^2]\leq 0$ or $c_s^2-c_n^2\geq 0$ within any region in the star, i.e.~at no point in the star can \textit{both} of these conditions be violated, however we find that such condition is automatically satisfied as long as $c_s^2\in[0,1]$.
Separately, we find two sufficient conditions for stability of high-frequency modes:
(i) $c_n^2\leq 0$ everywhere in the star, and (ii) the inequality of Eq.~\eqref{eq:condi}.
Then, if either of these sufficient conditions is satisfied, it suffices to check the stability of a finite number of low-frequency modes to determine if the star is radially stable to heat conductivity effects.

In the fast timescale, only the relaxation times of the MIS and BDNK contribute to first order. 
Thus, the two-timescale analysis separates the transport coefficients into two groups:
the ``classical'' coefficients $\zeta,\eta$, and $\kappa$ that appear only in the slow timescale,
and the relaxation times that only affect the fast timescale.
Nevertheless, the behavior of MIS and BDNK perturbations in the fast timescale is qualitatively different.
For a MIS fluid, we find that modes in the fast timescale are strictly decaying and quickly lead to the system relaxing to a perfect fluid.
This effect is more prominent for small viscosities.
Meanwhile, for a BDNK fluid, we find that the system has both decaying and oscillating behavior that is described by telegrapher equations.
We find the oscillation frequencies for a BDNK fluid [Eq.~\eqref{eq:fasteig}] and show that the oscillations can never lead to an instability.
Moreover, these frequencies allow us to find that the speed of propagation of the perturbations in the fast timescale is the speed of second sound $c_\tau=\sqrt{\tau_\mathscr P/\tau_\mathscr E}$ (up to factors of the metric), for which we obtain a causality constraint $\tau_\mathscr P<\tau_\mathscr E$.
This constraint is the same as the causality constraint of BDNK fluids, when assuming no bulk or shear viscosity and no heat conductivity in a flat background \cite{Bemfica2022}. 
This is to be expected since, at lowest order in the fast timescale, the equations depend only on the relaxation time.
Our results agree with the recent work \cite{RedondoYuste2024}, where the BDNK causality constraints were rederived by using neutron star perturbations for non-radial modes. 

We must note that the stability condition we obtained in the slow timescale differs from that previously obtained for neutron stars described by an Eckart fluid in \cite{Lindblom1983}. We expect the reason for this to be that we changed the heat flux vector for the Eckart and MIS fluids to resemble that of a BDNK fluid (see Sec.~\ref{ssec:rela}).
If we were to keep the original heat flux vector for those two models, then we expect the stability conditions to differ less drastically, and to recover the condition of \cite{Lindblom1983}.
Nevertheless, we establish an important result: the radial stability for small viscosity of a neutron star modeled with Eckart, BDNK and MIS fluids can \textit{only} differ in the heat flux.
That is, the degrees of freedom added by the BDNK and MIS models do not change the radial stability of neutron stars for small viscosity.
This is because the additional degrees of freedom only enter in the fast timescale part of the solution,
where all resulting modes have a positive real part, leading to quickly-decaying modes.
Therefore, the fast timescale --and the degrees of freedom associated with it-- does not affect stability.
We do note that the behavior in the fast timescale of the two models is expected to be different, with BDNK fluids leading to very fast oscillations that quickly decay away, while MIS fluids do not oscillate in the fast timescale.
However, since these modes are quickly decaying, we expect them to be difficult to observe.

Additional work on the stability of neutron stars in viscous theories remains to be done.
First, no analysis of the CFS instability has been carried out for causal viscous theories, with only \cite{Lindblom1983} addressing the instability for an Eckart fluid.
Whether the Newtonian results hold in general relativity --that the viscous instability and CFS instability counteract each other-- requires a treatment of the instability using a theory of fluids compatible with Einstein's theory, i.e.~the theory must be causal.
It would also be especially helpful to extend the results of this work to non-radial modes of rotating stars; specifically, it would be interesting to prove that the stability properties of more complicated viscous fluid models are the same as those of Eckart fluids.
If such a result is proven, then any conclusions from the linear stability analysis for an Eckart fluid would also apply for more complicated fluid theories.
This would include the result of \cite{Lindblom1983}, namely that viscosity can stabilize rotating stars against the CFS instability.
The study of non-radial modes of spherically-symmetric neutron stars within the BDNK framework has already been started with \cite{RedondoYuste2024}.
A stability analysis based on this work can be performed by calculating the eigenfrequencies, and possibly extend our results to non-radial modes by the use of singular perturbation theory.
However, to study the CFS instability, one would need to analyze the modes of \textit{rotating} neutron stars, and thus, the work of \cite{RedondoYuste2024} would need to be extended to a rotating background.
A full mode decomposition of rapidly rotating neutron stars has yet to be carried out even for a perfect fluid, but the perturbation equations have been derived in the Lorenz gauge and used to study the modes through numerical evolution in \cite{Kruger2020}.
It would be necessary to extend these equations for viscous fluids and decompose them into modes to understand how viscosity can dampen the CFS instability.
Other extensions of our work include studying the radial modes in other causal fluid models, such as the DNMR model, and others such as those described in \cite{Rocha2024}.
Finally, for the case of radial perturbations in a BDNK fluid, we leave for the future finding the eigenvalues through non-perturbative means, and extending the solution to higher orders in the perturbative expansion, preferably through the use of transseries or superasymptotic/hyperasymptotic series that address the apparent non-perturbative terms in the expansion.

\acknowledgments

We thank Jorge Noronha, Marcelo Disconzi and Abhishek Hegade for helpful discussions on the BDNK theory,
and Anand Balivada for discussion on the scaling of the radial variable. 
D. A. C. is thankful for the support from UIUC Graduate College and the Grainger College of Engineering, and from the Sloan Foundation.
N. Y. acknowledge support from the Simons Foundation through Award No. 896696, the NSF through Award No. PHY-2207650 and NASA through Grant No. 80NSSC22K0806.

\appendix

\section{Derivation of equations for radial perturbations}\label{app:der}

We now carry out the derivation for the perturbed Euler equation Eq.~\eqref{eq:pperp} in the radial direction.
The left-hand side is the same as for a perfect fluid, which we can write as \cite{Chandrasekhar1964,MTW,Caballero2024}\footnote{The equation can be written with either Eulerian or Lagrangian perturbations, both must be $0$.}
\begin{align}\label{eq:euleul}
    &(\delta\varepsilon+\delta p)a_r+(\varepsilon+p)\delta a_r+\delta(\Pi_r^{\lambda}\nabla_\lambda p) \nonumber\\
    =&
    e^{2(\Lambda+\Phi)}(\varepsilon+p)\ddot\xi+\partial_r\delta p-(\varepsilon+p)\partial_r\delta\Phi-(\delta\varepsilon+\delta p)\Phi' \ .
\end{align}
There are three new types of terms compared to the perfect-fluid equations:
1) terms that go like $\Delta s$ coming from $\delta p,\delta\varepsilon$ because the equation of state now depends on $s$ as given in Eqs.~\eqref{eq:devare}, \eqref{eq:devarp},
2) terms that go like $\Psi$ coming from $\delta p,\delta\varepsilon$ because we cannot assume anymore that $\Psi=0$ from Eq.~\eqref{eq:dtlam},
3) terms of $\delta\mathscr P,\delta\pi_r^r$ from the additional term in the right-hand side of Eq.~\eqref{eq:eferr}.
For the first two types of new terms, we see they come from
\begin{align}
    \delta\varepsilon&=-\frac{\varepsilon+p}{c_s^2}\Phi'\xi-\frac{(\varepsilon+p)e^{-\Phi}}{r^2}\partial_r(r^2e^\Phi\xi)\\
    &\quad+(\varepsilon+p)\Psi+nT\Delta s  \ ,\\
    \delta p&=-(\varepsilon+p)\Phi'\xi-\frac{\gamma pe^{-\Phi}}{r^2}\partial_r(r^2e^\Phi\xi)\\
    &\quad+\gamma p\Psi+nTc_n^2\Delta s \ ,
\end{align}
in particular, only the terms in the second line of each are new.
We then see that the new terms with $\Psi$ in Eq.~\eqref{eq:euleul} after multiplying by $e^{\Lambda-2\Phi}$ are
\begin{align}
    &e^{\Lambda-2\Phi}\big[\partial_r(\gamma p\Psi)-\left((\varepsilon+p)+\gamma p\right)\Phi'\Psi \nonumber\\
    &\quad\quad +4\pi re^{2\Lambda}(\varepsilon+p)\gamma p \Psi+(2\Phi'-\tfrac1r)(\varepsilon+p)\Psi\big] \nonumber\\
    =&\partial_r\left(\gamma pe^{\Lambda-2\Phi}\Psi\right)+(\varepsilon+p)\left[\Phi'-\tfrac1r\right]e^{\Lambda-2\Phi}\Psi \nonumber\\
    &\ -\left[(\Lambda'-2\Phi') -4\pi re^{2\Lambda}(\varepsilon+p)+\Phi'\right]e^{\Lambda-2\Phi}\gamma p\Psi\ ,
\end{align}
then, from Eq.~\eqref{eq:rela}, the second line cancels out and we get the negative of the second line of Eq.~\eqref{eq:bigg2}.
Notice that the terms in the second line come from the dependence of $\partial_r\delta\Phi$ on $\delta\Lambda$ (and hence $\Psi$) in Eq.~\eqref{eq:eferr}, and on $\delta p$ in the same equation.
We can do the same for the terms that have $\Delta s$ in them in Eq.~\eqref{eq:euleul} will be of the form
\begin{align}
    &e^{\Lambda-2\Phi}\left[\partial_r(c_n^2nT\Delta s)-(nT+c_n^2nT)\Phi'\Delta s\right]\\
    &\quad +4\pi re^{2\Lambda}(\varepsilon+p)nT\Delta s\\
    =&\partial_r(e^{\Lambda-2\Phi} c_n^2nT\Delta s)-\Phi'nT\Delta s \ ,
\end{align}
where we once again made use of Eq.~\eqref{eq:rela} to simplify, and see the we once again obtain the negative of the third line of Eq.~\eqref{eq:bigg2}.
We now proceed to evaluate the right-hand side of Eq.~\eqref{eq:pperp}.
Since the left-hand side as given in Eq.~\eqref{eq:euleul} is the same whether we take Eulerian or Lagrangian perturbations, we can evaluate the right-hand side of Eq.~\eqref{eq:pperp} using Lagrangian perturbations even though we evaluated the left-hand side using Eulerian perturbations.
We quickly evaluate the right-hand side to be
\begin{align}
    &-(\Delta\mathscr E+\Delta \mathscr P)a_r-\Pi_r^{\lambda}\nabla_\lambda\Delta\mathscr P-\Pi_r^{\lambda}\nabla_\nu(\Delta\pi^\nu_\lambda) \nonumber\\
    &\quad\quad-u^\nu\nabla_\nu\Delta\mathscr Q_r \nonumber\\
    =&(\Delta\mathscr E+\Delta\mathscr P)\Phi'-\partial_r\Delta\mathscr P-\partial_r\Delta\pi_r^r+(\Phi'-\tfrac3r)\Delta\pi_r^r\nonumber \\
    &\quad\quad-e^\Phi\partial_t\Delta\mathscr Q_r \ ,
\end{align}
where we have used that $\Delta\pi_2^2+\Delta\pi_3^3=-\Delta\pi_r^r$ to simplify the expression involving the Christoffel symbols.
The only other contributions from viscous terms that show up for the equation will come from $(\varepsilon+p)\partial_r\delta\Phi$ on the left-hand side of Eq.~\eqref{eq:euleul}.
Seeing that $\delta\mathscr P=\Delta\mathscr P,\delta\pi_r^r=\Delta\pi_r^r$, 
we see that the contributions to Eq.~\eqref{eq:pperp} from all viscous components (after multiplying by $e^{\Lambda-2\Phi}$) are
\begin{align}
    &e^{\Lambda-2\Phi}\big[\Delta\mathscr E\Phi'+\Phi'(\Delta\mathscr P+\Delta\pi_r^r)-\partial_r(\Delta\mathscr P+\Delta\pi_r^r)-\tfrac3r\Delta\pi_r^r \nonumber \\
    &\quad\quad -e^\Phi\partial_t\Delta\mathscr Q_r+4\pi re^{2\Lambda}(\varepsilon+p)(\Delta\mathscr P+\Delta\pi_r^r)\big] \nonumber\\
    &=-\partial_r\left[e^{\Lambda-2\Phi}(\Delta\mathscr P+\Delta\pi_r^r)\right]-\tfrac3re^{\Lambda-2\Phi}\Delta\pi_r^r \nonumber\\
    &\quad -e^{\Lambda-\Phi}\partial_t\Delta\mathscr Q_r+e^{\Lambda-2\Phi}\Delta\mathscr E\Phi' \ ,
\end{align}
where we once again made use of Eq.~\eqref{eq:rela} to simplify terms.
Therefore, Eq.~\eqref{eq:pperp} can be written as
\begin{widetext}
    \begin{align}\label{eq:biggo}
    W(r)\partial_t\dot\Xi=&\partial_r(P(r)\partial_r\Xi)-Q(r)\Xi \nonumber\\
    &+\partial_r\left(e^{\Lambda-2\Phi}\gamma p\Psi\right)+(\varepsilon+p)e^{\Lambda-2\Phi}\Psi\left[\Phi'-\frac1r\right] \nonumber\\
    &-\partial_r\left(e^{\Lambda-2\Phi}c_n^2nT\Delta s\right)+\Phi' e^{\Lambda-2\Phi}nT \Delta s \nonumber\\
    &-e^{\Lambda-\Phi}\partial_t\Delta\mathscr Q_r-\partial_r\left[e^{\Lambda-2\Phi}(\Delta\mathscr P+\Delta\pi_r^r)\right]-\frac3re^{\Lambda-2\Phi}\Delta\pi_r^r +e^{\Lambda-2\Phi}\Delta\mathscr E\Phi'\ .
\end{align}
\end{widetext}

Notice that this expression is general for any relativistic viscous theory which makes use of the decomposition Eq.~\eqref{eq:corr} that has $J^\mu_{\rm NPF}=0$.
To obtain the expression for each of the fluids, we must evaluate the different values of the perturbations of the viscous fluxes.
For example, for the Eckart fluid, it suffices to evaluate $\Delta\mathscr E=0,\Delta\mathscr P=-\zeta\Delta\Theta,\Delta\pi_r^r=-2\eta\Delta\sigma_r^r$ and using Eq.~\eqref{eq:velperts} to obtain expressions for $\Delta\Theta,\Delta\sigma_r^r$.
Notice that $\Delta\Theta$ can be written in terms of the $\Xi,\Psi$ variables as
\begin{align}
    \Delta\Theta&=e^\Phi\left[\partial_t\Psi-\frac{4\pi}{r} e^{2\Lambda-\Phi}\partial_t\Xi+(\tfrac2r+\Lambda'+\partial_r)\partial_t\left(e^{-\Phi}\Xi/r^2\right)\right] \nonumber\\
    &=e^\Phi\left[-4\pi re^{-\Phi}\Delta\mathscr Q_r+\frac{e^{-\Phi}}{r^2}\partial_r\dot\Xi\right] \ ,
\end{align}
where we once again make use of Eq.~\eqref{eq:rela} plus Eq.~\eqref{eq:psi} to remove the $\partial_t\Psi$ term.
Substituting $\Delta\mathscr E=0$, this value of $\Delta\Theta$ and $\Delta\pi_r^r$ from Eq.~\eqref{eq:velperts} into Eq.~\eqref{eq:biggo}, results in Eq.~\eqref{eq:bigg2} for the Eckart fluid.
To obtain Eq.~\eqref{eq:bigg3} for the MIS fluid, we first solve for $\partial_t\Delta\mathscr Q_r$ in Eq.~\eqref{eq:misq}, then we take $\Delta\mathscr E=0$ and substitute the result into Eq.~\eqref{eq:biggo}. 
Finally, for the BDNK fluid, substituting $\Delta\mathscr P=-\zeta\Delta\Theta+\frac{\tau_\mathscr P}{\tau_\mathscr E}\Delta\mathscr E$, and $\Delta\pi_r^r=2\Delta\sigma_r^r$ with the same expressions for $\Delta\Theta,\Delta\sigma_r^r$ as for the Eckart fluid, together with Eq.~\eqref{eq:dq} for $\partial_t\Delta\mathscr Q_r$ and a lot of simplifications, we obtain Eq.~\eqref{eq:biggg}.

\section{Symmetric and Anti-Symmetric parts of a fourth order operator}\label{app:fourth}

In this appendix, we show how to write a fourth-order differential operator into a symmetric and antisymmetric part and then derive what these are for $\mathbb G$ of Sec.~\ref{ssec:smalleck}.
First, let us show how this works for a second-order operator $\mathbb L_2$ which we can generally write
\begin{equation}\label{eq:ord2}
    \mathbb L_2\phi=P(r)\partial_r^2\phi+T(r)\partial_r\phi+S(r)\phi \ .
\end{equation}
The operator can alternatively be written as
\begin{align}
    \mathbb L_2\phi=&\partial_r\left(P(r)\partial_r\phi\right)+\left[S(r)-\tfrac12T'(r)+\tfrac12P''(r)\right]\phi \nonumber\\
    &+\frac12\left[\left(T(r)-P'(r)\right)\partial_r\phi+\partial_r\left(\left(T(r)-P'(r)\right)\phi\right)\right] \ ,
\end{align}
which can be checked by simply expanding all derivatives.
Defining $Q(r)=S(r)-\tfrac12T(r)+\tfrac12P''(r)$ and $O(r)=T(r)-P'(r)$ then the first line of the previous equation is clearly written in Sturm-Liouville form.
We can then define each line as its own operator, that is
\begin{subequations}
    \begin{align}
        \mathbb S_2\phi&=\partial_r\left(P(r)\partial_r\phi\right)+Q(r)\phi \ ,\\
        \mathbb A_2\phi&=\frac12\left[O(r)\partial_r\phi+\partial_r(O(r)\phi)\right] \ .
    \end{align}
\end{subequations}
We now check for symmetry within the space of square integrable functions on $(0,R)$\footnote{We will make the calculations in the unweighted space. Symmetry in the unweighted space for $\mathbb S_2$ implies that $W(r)^{-1}\mathbb S_2$ is symmetric in the weighted space.}.
As usual, we find
\begin{align}\label{eq:bcs2}
    \int_0^R dr\, \eta^*\mathbb S_2\phi=&P(r)\eta^*\partial_r\phi\eval_0^R-\int_0^Rdr\, P(r)\partial_r\eta^*\partial_r\phi \nonumber\\
    &+\int_0^Rdr\, Q(r)\eta^*\phi \ ,
\end{align}
which says that as long as the boundary term is $0$, the operator is symmetric.
We can do the same for $\mathbb A_2$ where an integration by parts leads to
\begin{align}\label{eq:bca2}
    \int_0^Rdr\, \eta^*\mathbb A_2\phi=&\frac12O(r)\eta^*\phi^*\eval_0^R \nonumber\\
    &+\frac12\int_0^Rdr\, O(r)\left[\eta^*\partial_r\phi-\partial_r\eta^*\phi\right] \ ,
\end{align}
which once again says that as long as the boundary term is $0$, $\mathbb A_2$ will be anti-symmetric.

We can now use this for the idea behind a fourth order operator.
We can generally write a fourth order operator by
\begin{align}
    \mathbb L_4\phi=&\partial_r^2\left(L(r)\partial_r^2\phi\right)-\partial_r(M(r)\partial_r\phi)+N(r)\phi \nonumber\\
    &+\frac12\left[E(r)\partial_r^3\phi+\partial_r^3(E(r)\phi)\right] \nonumber\\
    &+\frac12\left[F(r)\partial_r\phi+\partial_r(F(r)\phi)\right] \ .
\end{align}
Then we define $\mathbb S_4$ from the first line and $\mathbb A_4$ from the second and third lines.
We now check the integrals for the operators same as before
\begin{align}
    \int_0^Rdr\, \eta^*\mathbb S_4\phi=&\bigg[\eta^*\partial_r(L\partial_r^2\phi)-L\partial_r\eta^*\partial_r^2\phi-M\eta^*\partial_r\phi\bigg]_0^R \nonumber\\
    +&\int_0^Rdr\, \left[L\partial_r^2\eta^*\partial_r^2\phi+ M\partial_r\eta^*\partial_r\phi+ N\eta^*\phi\right] \ , \label{eq:symm4}\\
    \int_0^R\eta^*\mathbb A_4\phi=&\frac12\bigg[\eta^*\partial_r^2(E\phi)-\partial_r\eta^*\partial_r(E\phi)\nonumber\\
    &\quad+ \partial_r^2\eta^*E\phi+\eta^*F\phi\bigg]_0^R\nonumber\\
    &+\frac12\int_0^Rdr\,\big[ E\left(\eta^*\partial_r^3\phi-\partial_r^3\eta^*\phi\right)\nonumber\\
    &\quad\quad\quad+F\left(\eta^*\partial_r\phi-\partial_r\eta^*\phi\right)\big] \label{eq:asymm4} \ .
\end{align}
Then, as long as the boundary terms in the first line of Eq.~\eqref{eq:symm4} are $0$, $\mathbb S_4$ will be symmetric.
Similarly, as long the boundary terms in the first two lines of Eq.~\eqref{eq:asymm4} are $0$, $\mathbb A_4$ will be anti-symmetric.
Given that the boundary terms are $0$, the expected values for said operators are then
\begin{align}
    \braket{\phi}{\mathbb S_4\phi}&=\int_0^Rdr\, L|\partial_r^2\phi|^2+M|\partial_r\phi|^2+N|\phi|^2\ ,\\
    \braket{\phi}{\mathbb A_4\phi}&=i\int_0^Rdr\, E(r)\Im(\phi^*\partial_r^3\phi)+F(r)\Im(\phi^*\partial_r\phi) \ . \label{eq:eva4}
\end{align}

We now proceed to check that the boundary terms are indeed $0$ for the symmetric and antisymmetric operators that come out of $\mathbb G$ and the heat conductivity part of $\mathbb F$ are indeed $0$.
For this, we require writing out $L,M,N,E,F$ for $\mathbb G$ and $K,H,O$ for $\mathbb F$ when $\zeta=0,\eta=0$.
We first do this for $\mathbb F$, which can be written easily in the form Eq.~\eqref{eq:ord2} as
\begin{align}
    \mathbb F\phi=&\partial_r\left[e^{\Lambda-\Phi}\kappa \left(\frac{\varepsilon+p}{n}\right)^2(c_s^2-c_n^2)\frac{e^{-\Phi}}{r^2}\partial_r\phi\right] \nonumber\\
    &-\partial_r\left(e^{\Lambda-\Phi}\kappa T\frac{\varepsilon+p}{n}\right)\frac{\varepsilon+p}{nT}(c_s^2-c_n^2)\frac{e^{-\Phi}}{r^2}\partial_r\phi \ .
\end{align}
Then, $K$ is the term in the first parentheses and $O$ is the coefficient of $\partial_r\phi$ in the second line.
The final coefficient, will be given by $H=-\frac12\partial_rO$.
Therefore, we have found
\begin{align}
    K&=e^{\Lambda-\Phi}\kappa \left(\frac{\varepsilon+p}{n}\right)^2(c_s^2-c_n^2)\frac{e^{-\Phi}}{r^2} \ ,\\
    O&=-\partial_r\left(e^{\Lambda-\Phi}\kappa T\frac{\varepsilon+p}{n}\right)\frac{\varepsilon+p}{nT}(c_s^2-c_n^2)\frac{e^{-\Phi}}{r^2} \ ,\\
    H&=\frac12\partial_r\left[\partial_r\left(e^{\Lambda-\Phi}\kappa T\frac{\varepsilon+p}{n}\right)\frac{\varepsilon+p}{nT}(c_s^2-c_n^2)\frac{e^{-\Phi}}{r^2}\right] \ .
\end{align}
Comparing this with the boundary term from Eq.~\eqref{eq:bcs2}, at $0$ we know that for eigenfunctions $\eta,\phi\sim r^3$ while $K\sim r^{-2}$, and hence $\eta^*K\partial_r\phi\sim r$ at $0$.
Meanwhile, at $R$, we must impose that
\begin{equation}\label{eq:cond}
    \lim_{r\to R}\frac{\kappa(c_s^2-c_n^2)\partial_r\phi}{r^2}=0
\end{equation}
This is similar to how we had to impose for $(\zeta+\tfrac43\eta)$ to follow a similar behavior.
Moreover, since $\gamma p\partial_r\phi\to 0$ and $\gamma p\to 0$ at the boundary, then, as long as $\kappa(c_s^2-c_n^2)\to0$ at least as fast $\gamma p\to 0$ at the boundary, the boundary term will be $0$.
We note that even though $(\varepsilon+p)\to 0$ at the boundary, the ratio $(\varepsilon+p)/n$ does not necessarily go to $0$ since $n\to 0$ at the boundary.
Now, for the boundary term in Eq.~\eqref{eq:bca2}, the boundary term at the origin is $0$ by a similar reasoning as for that of the symmetric operator.
For the boundary at $R$, the function $O$ has two terms, one with $\partial_r\kappa$ and another which just multiplies by $\kappa$.
For the latter, since $\kappa(c_s^2-c_n^2)\to 0$ at $R$, and any eigenfunction $\phi$ must be finite at $R$, it follows that such term is $0$.
This leaves us with the last term being
\begin{equation}\label{eq:limiO}
    \frac12 O\eta^*\phi^*\eval_0^R=-\frac12\left(\frac{\varepsilon+p}{n}\right)^2\frac{e^{\Lambda-2\Phi}\partial_r\kappa(c_s^2-c_n^2)}{r^2}\eta^*\phi\eval_R \ .
\end{equation}
Since $\phi,\eta$ are finite and not necessarily $0$ at $R$, and as mentioned before $(\varepsilon+p)/n$ does not need vanish at the boundary, we will have to require that $\partial_r\kappa(c_s^2-c_n^2)\to 0$ at the boundary.

Checking the boundary terms for $\mathbb G$ will be more complicated but it follows the same idea.
The functions beyond $L$ are complicated to find outside of the happy coincidence that $N=0$.
As stated in the body, $L$ is given by
\begin{equation}
    L=-\kappa\left(\frac{\varepsilon+p}{n}\right)^2\frac{e^{-\Lambda-4\Phi}}{r^2}c_n^2\left[c_s^2-c_n^2\right] \ .
\end{equation}
We first check the boundary conditions at $0$.
By the fact that $\eta\sim r^3,\phi\sim r^3$, then $L\partial_r\eta^*\partial_r^2\phi\sim r$ at $0$.
The term $\eta^*L\partial_r^2\phi\sim r$ at $0$ too.
Notice that this is the same behavior at $0$ as the term involving $K$ in the above.
We still need to check $\eta^*\partial_rL\partial_r^2\phi$ at the origin.
Explicitly, this is
\begin{equation}
    \eta^*\partial_rL\partial_r^2\phi\eval_0\sim r^4\partial_r\left(\frac{1}{r^2}\right)\sim r \ ,
\end{equation}
where we used that $\phi,\eta\sim r^3,L\sim r^{-2}$.
Now, for the boundary term at $R$, from Eq.~\eqref{eq:cond} the term $L\partial_r\eta^*$ must go to $0$ at the boundary of the star.
Moreover, since $\eta^*$ must be finite at the boundary, $L\eta^*\partial_r^2\phi$ also goes to $0$.
This leaves only the $\eta^*\partial_rL\partial_r^2\phi$ term.
We do the same as we did for $O$: we have a term that multiplies $\kappa$ and another that multiplies $\partial_r\kappa$.
The term that multiplies $\kappa$ will go to zero due to the condition $\kappa(c_s^2-c_n^2)\to 0$, while that of $\partial_r\kappa$ will be similar to that of Eq.~\eqref{eq:limiO}, requiring a condition of $\partial_r\kappa(c_s^2-c_n^2)\to 0$.

For the remaining boundary terms, the functions $E,M,F$ can become quite complicated.
However, it is easy to see that $E\sim \partial_rL,M\sim\partial_r^2L,F\sim\partial_r^3L$.
Then, boundary terms at $0$ such as $M\eta^*\partial_r\phi\sim\eta^*\partial_r^2L\partial_r\phi\sim r^5\partial_r^2(r^{-2})\sim r$ near $0$.
This will hold for all boundary terms near the origin, and thus, all boundary terms near the origin go to $0$.
At the boundary of the star, the conditions become more complicated.
However, we will obtain similar conditions where derivatives thermodynamic quantities such as $\partial_rc_n^2,\partial_rc_s^2,\partial_r\kappa$ and higher order derivatives must go to $0$ at the boundary.
If such is the case, then since $\phi,\eta^*$ and their derivatives are finite at the boundary, then the boundary term will go to $0$.

Since we have established --albeit without directly calculating-- that the boundary terms go to zero, then $\mathbb G$ and $\mathbb F$ can be decomposed into the symmetric and anti-symmetric part.
It then follows that the only contribution to the stability of comes from their symmetric parts and Eq.~\eqref{eq:heatint} holds.
As for calculating the new frequencies, from Eq.~\eqref{eq:evcorr} we find
\begin{equation}
    \Re(\omega^{(1)})=\frac12\frac{|\braket{\phi_j}{\mathbb A_2\phi_j}|}{\braket{\phi_j}}-\frac12\frac{1}{\left(\omega^{(0)}\right)^2}\frac{|\braket{\phi_j}{\mathbb A_4\phi_j}|}{\braket{\phi_j}} \ ,
\end{equation}
which from Eq.~\eqref{eq:eva4} (and its equivalent for the second-order operator) we find that
\begin{align}
    2\Re(\omega^{(1)})=&\frac{\int_0^Rdr\, O(r)\Im\left((\phi_j^{(0)})^*\partial_r\phi_j^{(0)}\right)}{\int_0^Rdr\, W|\phi_j^{(0)}|^2} \nonumber\\
    -&\frac{\int_0^Rdr\, E\Im\left((\phi_j^{(0)})^*\partial_r^3\phi_j^{(0)}\right)+F\Im\left((\phi_j^{(0)})^*\partial_r\phi_j^{(0)}\right)}{\int_0^Rdr\, P|\partial_r\phi_j^{(0)}|^2+Q|\phi_j^{(0)}|^2} \ .
\end{align}
We refrain from explicitly writing out $E,F,M$ due to their large size.


\bibliography{Paper_Radial/references}

\end{document}